\pdfoutput=1
\documentclass[PAPER, atlasdraft=false, UKenglish,cernpreprint, texmf, orcidlogo]{atlasdoc}
\usepackage[subfigure]{atlaspackage}
\usepackage{atlasbiblatex}

\usepackage{atlasphysics}
\graphicspath{{logos/}{figures/}}

\usepackage{ANA-JETM-2023-08-PAPER-defs}
\AtlasTitle{Using pile-up collisions as an abundant source of low-energy hadronic physics processes in ATLAS and an extraction of the jet energy resolution}

\AtlasAbstract{%
During the 2015--2018 data-taking period, the Large Hadron Collider delivered proton--proton bunch crossings at a centre-of-mass energy of 13\,TeV to the ATLAS experiment at a rate of roughly 30\,MHz, where each bunch crossing contained an average of 34 independent inelastic proton--proton collisions.
The ATLAS trigger system selected roughly 1\,kHz of these bunch crossings to be recorded to disk.
Offline algorithms then identify one of the recorded collisions as the collision of interest for subsequent data analysis, and the remaining collisions are referred to as pile-up.

Pile-up collisions represent a trigger-unbiased dataset, which is evaluated to have an integrated luminosity of 1.33\,pb$^{-1}$ in 2015--2018.
This is small compared with the normal trigger-based ATLAS dataset, but when combined with vertex-by-vertex jet reconstruction it provides up to 50 times more dijet events than the conventional single-jet-trigger-based approach, and does so without adding any additional cost or requirements on the trigger system, readout, or storage.
The pile-up dataset is validated through comparisons with a special trigger-unbiased dataset recorded by ATLAS, and its utility is demonstrated by means of a measurement of the jet energy resolution in dijet events, where the statistical uncertainty is significantly reduced for jet transverse momenta below 65\,GeV.
}

\AtlasRefCode{JETM-2023-08}

\PreprintIdNumber{CERN-EP-2024-192}

\AtlasJournalRef{\JHEP{} 12 (2024) 032}
\AtlasDOI{10.1007/JHEP12(2024)032}

\AtlasCoverEgroupAnalysisTeam{atlas-jetm-2023-08-analysis-team@cern.ch}

\hypersetup{pdftitle={ATLAS document},pdfauthor={The ATLAS Collaboration}}

\begin{document}

\maketitle

\section{Introduction}
\label{sec:intro}

%

The Large Hadron Collider (LHC)~\cite{Evans:2008zzb} accelerates counter-rotating beams of protons to unprecedented energies in a controlled laboratory environment.
After reaching the desired energy, the beams are brought into collision at four different Interaction Points (IPs) around the ring of the LHC\@.
The instances at which the beams are brought into collision are referred to as bunch crossings, and can result in proton--proton (\ppp{}) collisions.
During Run~2 of the LHC, from 2015--2018, the \ppp{} collision centre-of-mass energy was 13\,\TeV{}, and the typical LHC filled bunch crossing rate was roughly 30\,MHz.

The ATLAS detector~\cite{PERF-2007-01}, located at one of the LHC IPs, analyses each of the \ppp{} collisions delivered by the LHC\@.
Numerous particles are produced in these collisions, and those that are incident upon the detector create signals.
These signals are processed during data-taking by the ATLAS trigger and data acquisition system, which, during Run~2, selected an average of 1\,kHz of bunch crossings to be recorded~\cite{TRIG-2016-01}.
Each of these bunch crossings contains an average of roughly 34 inelastic \ppp{} collisions.
In this paper, the term \textit{recorded bunch crossing} (\rbc{}) is used to refer to the full activity of a bunch crossing selected by the trigger and data acquisition system, including the full set of detector signals from all of the produced particles.
The term \textit{event} is instead used to refer to the detector signals associated with a physics process generated by a single \ppp{} collision.

Despite the presence of numerous collisions in a single bunch crossing, in most cases the decision to record the bunch crossing is triggered by a single collision; offline algorithms subsequently identify the collision of interest, and the remaining collisions are referred to as pile-up.
The typical trigger-based data analysis therefore focuses on the event of interest, and discards the pile-up collisions, resulting in an interpretation of one event of interest per \rbc{}.
The LHC Run~2 dataset, as recorded by the ATLAS detector, has made use of this trigger-based methodology to support a diverse research programme, including precision measurements of Standard Model (SM) processes and searches for new phenomena Beyond the Standard Model (BSM).
This approach works very well when the \textit{process of interest} (\poi{}) has a clear trigger signature, and is sufficiently rare that it is feasible to record a reasonable fraction of the \poi{} within the constraints of the overall 1\,kHz recording rate.

In this paper, an innovative pile-up-based methodology is developed, which opens up new possibilities for data analysis in cases where it is not possible to trigger on the \poi{}, or in cases where the \poi{} is so abundant that recording a sufficiently large dataset would interfere with overall recording rate constraints, or even overwhelm the storage system.
This new approach relies on the independence of each \ppp{} collision within a single \rbc{}: by removing the collision(s) that caused the event to be recorded, the remaining collisions form a trigger-unbiased\footnote{This follows the ATLAS definition of a zero-bias trigger, which refers to collisions that take place within filled bunch crossings.} dataset, which can be used for any physics analysis of interest.
Moreover, this dataset is created by reinterpreting already recorded data, thus it comes without any additional demands on the trigger, data acquisition, or storage systems.
The concept behind this approach has parallels to that proposed in Ref.~\cite{Nachman:2016nes}.

In order to develop such a pile-up-based approach, it is necessary to make extensive usage of charged-particle tracks as reconstructed by the ATLAS inner detector, and the vertices from which multiple tracks originate.
When these vertices are consistent with the spatial region in which the beams overlap, they are referred to as \textit{primary vertices} (\pvs{})~\cite{PERF-2015-01,ATL-PHYS-PUB-2015-026}.
The large majority of the collisions can either be mapped to a single \pv{}, or are not visible to the ATLAS inner detector,\footnote{The main reasons for which a \pv{} is not visible by the ATLAS inner detector are either that the charged-particle tracks are outside of the detector acceptance, the interactions are so low-energy that the resulting tracks are below the reconstruction thresholds, or the physics process has minimal charged activity and thus does not produce tracks.} thus there is primarily a one-to-one mapping between observed \pvs{} and distinct \ppp{} collisions~\cite{IDTR-2018-006}.
This mapping enables the efficient removal of the triggering collision, by identifying the \textit{triggering primary vertex} (\tpv{}), as well as the recovery and usage of each of the pile-up collisions, by reconstructing each of the \textit{pile-up primary vertices} (\ppvs{}).

The majority of the \ppvs{} that produce physics objects with sufficient transverse momentum (\pT{}) to satisfy analysis selection requirements originate as the result of $2\to2$ quark and/or gluon scattering processes.
However, quarks and gluons are not detector-stable particles, rather they form collimated sets of particles through fragmentation and hadronisation processes, with the end result being referred to as a shower.
These showers are observed as localised regions of signals within the detector, and are typically grouped together into jets; ideally, each jet contains the full shower, and thus can be used as a proxy for the original quark or gluon.
The most abundant signature from pile-up collisions, with sufficient momentum transfer to produce substantial detector signals, is therefore the production of two or more jets, where the jets are consistent with originating from a \ppv{}.

\begin{figure}
\centering
\includegraphics[width=\textwidth]{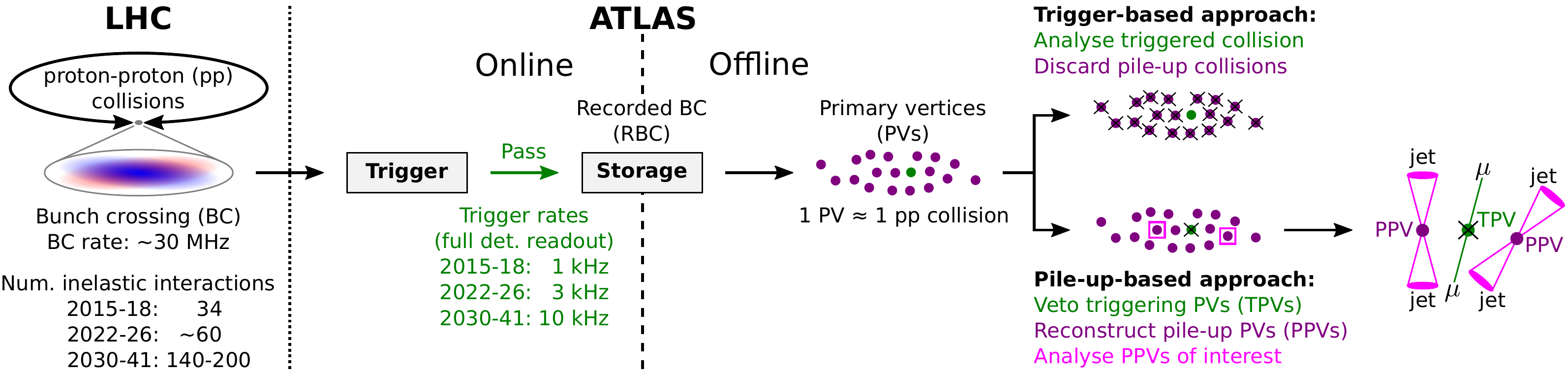}
\caption{
A schematic overview of the pile-up-based data analysis methodology.
}
\label{fig:overview}
\end{figure}

With this, all of the necessary pieces for the pile-up-based methodology are in place, as schematically depicted in \Figure{fig:overview}.
The potential of this approach depends strongly on the ATLAS trigger rate, which defines the \rbc{} rate, in comparison to the LHC filled bunch crossing rate.
The ratio of these two numbers approximates the effective integrated luminosity of the approach as a fraction of the total integrated luminosity: the ATLAS trigger system processes, and selects from, every \ppp{} collision of every LHC filled bunch crossing, while the analysis of \ppvs{} is only possible for \ppp{} collisions belonging to \rbcs{} written to disk as selected by the trigger system.

The extent to which it is possible to study low-energy hadronic physics by reconstructing pile-up collisions is linked to the ATLAS trigger rate, which defines how much data is available for subsequent offline \ppv{} reconstruction.
The ATLAS trigger rate has increased from 1\,kHz in 2015~\cite{TRIG-2016-01} to 3\,kHz in 2022~\cite{TRIG-2022-01}, and is planned to grow to 10\,kHz in 2030~\cite{ATLAS-TDR-29-difftitle}; each of these increases in trigger rate will enhance the utility of the pile-up dataset.
This augmented sensitivity is further enhanced by the increase in the average number of inelastic \ppp{} interactions per \bc{}, rising from 34 in Run~2 to about 60 in 2022--2026, and expected to reach 140--200 in 2030 and beyond.

This paper presents an overview of the pile-up-based methodology, starting with an overview of the ATLAS detector in \Section{sec:detector} and the data and simulated samples used for these results in \Section{sec:datasamples}.
The integrated luminosity of the pile-up dataset is evaluated in \Section{sec:lumi}, while the procedures used to reconstruct physics objects from individual \ppp{} collisions are detailed in \Section{sec:reco}.
Aspects related to the reconstruction and interpretation of entire bunch crossings, including how to define which pile-up collisions can be used, are discussed in \Section{sec:eventint}.
The pile-up dataset is used to extract the jet energy resolution in dijet events and is compared with previous results, providing a concrete example of the benefits of the pile-up-based methodology for physics analysis; this is documented in \Section{sec:JER}.
The paper concludes in \Section{sec:conclusions}.


%
\section{ATLAS detector}
\label{sec:detector}

%

The ATLAS detector~\cite{PERF-2007-01} at the LHC covers nearly the entire solid angle around the collision point.\footnote{\AtlasCoordFootnote}
It consists of an inner tracking detector surrounded by a thin superconducting solenoid, electromagnetic and hadron calorimeters,
and a muon spectrometer incorporating three large superconducting air-core toroidal magnets.

The inner-detector system (ID) is immersed in a \qty{2}{\tesla} axial magnetic field
and provides charged-particle tracking in the range of \(|\eta| < 2.5\).
The high-granularity silicon pixel detector covers the vertex region and typically provides four measurements per track,
the first hit normally being in the insertable B-layer (IBL) installed before Run~2~\cite{ATLAS-TDR-19,PIX-2018-001}.
It is followed by the silicon microstrip tracker (SCT), which usually provides eight measurements per track.
These silicon detectors are complemented by the transition radiation tracker (TRT),
which enables radially extended track reconstruction up to \(|\eta| = 2.0\).
The TRT also provides electron identification information
based on the fraction of hits (typically 30 in total) above a higher energy-deposit threshold corresponding to transition radiation.

The calorimeter system covers the pseudorapidity range of \(|\eta| < 4.9\).
Within the region \(|\eta|< 3.2\), electromagnetic calorimetry is provided by barrel and
endcap high-granularity lead/liquid-argon (LAr) calorimeters,
with an additional thin LAr presampler covering \(|\eta| < 1.8\)
to correct for energy loss in material upstream of the calorimeters.
Hadron calorimetry is provided by the steel/scintillator-tile calorimeter,
segmented into three barrel structures within \(|\eta| < 1.7\), and two copper/LAr hadron endcap calorimeters.
The solid angle coverage is completed with forward copper/LAr and tungsten/LAr calorimeter modules
optimised for electromagnetic and hadronic energy measurements respectively. %

The muon spectrometer (MS) comprises separate trigger and
high-precision tracking chambers measuring the deflection of muons in a magnetic field generated by the superconducting air-core toroidal magnets.
The field integral of the toroids ranges between \num{2.0} and \qty{6.0}{\tesla\metre}
across most of the detector.
Three layers of precision chambers, each consisting of layers of monitored drift tubes, cover the region \(|\eta| < 2.7\),
complemented by cathode-strip chambers in the forward region, where the background is highest.
The muon trigger system covers the range of \(|\eta| < 2.4\) with resistive-plate chambers in the barrel, and thin-gap chambers in the endcap regions.

The Trigger and Data Acquisition (TDAQ) system is responsible for identifying interesting physics events from the detector readout and deciding whether
they should be stored for offline analysis. These decisions are carried out by a two-stage trigger strategy implemented in the real-time data path~\cite{TRIG-2016-01}.

The Level-1 (L1) trigger is a hardware-based system using custom electronics that performs early rejection based on reduced
granularity information from the calorimeters and the muon spectrometer. The L1 Calorimeter (L1Calo) sub-system takes signals
from the calorimeters as input in order to reconstruct electron, photon, $\tau$-lepton and jet candidates. It then produces
decisions based on candidate multiplicity over programmable thresholds. Additionally, it also provides trigger decisions based on
global event properties such as missing transverse momentum in the collision. The L1Muon sub-system uses hits from the MS trigger chambers to reconstruct
the momentum of muon candidates and provide decisions based on their multiplicity over thresholds. The L1 Topological (L1Topo) sub-system
combines information from the previous two to formulate decisions based on topological properties of the candidates. The final L1 decision for each bunch crossing is
issued in the Central Trigger Processor (CTP), where the information from L1Calo, L1Muon and L1Topo is combined. This first stage
manages an event rate reduction from the initial LHC collisions at an average of roughly 30~MHz down to 100~kHz. The selected events are
propagated to the second stage, together with the identified Regions of Interest (RoIs) in the detector, for further processing.

The High-Level Trigger (HLT) is the second step in the ATLAS trigger chain. It is a software-based system that executes sequences of
algorithms to issue the final decision to store or reject events. A series of dedicated fast trigger algorithms is executed first
for early rejection, followed by reconstruction algorithms similar to the ones used offline for precise object reconstruction. These
sequences generally include running feature-extraction algorithms on data-fragments within the RoIs identified by
the L1 system. A hypothesis algorithm is then executed based on the reconstructed features to determine whether the configured
trigger condition is fulfilled. In the case of global event features, such as missing transverse momentum in the event, full detector information is used.
The final physics output rate of the HLT was on average 1.0 (1.2)~kHz, with an average physics throughput of 1.0 (1.2)~GB/s in 2015--2017 (2018).

A software suite~\cite{SOFT-2022-02} is used in data simulation, in the reconstruction and analysis of real and simulated data, in detector operations, and in the trigger and data acquisition systems of the experiment.


%
%
\section{Recorded and simulated data}
\label{sec:datasamples}

%

The results presented in this paper make use of \ppp{} collisions at centre-of-mass energy of $\sqrt{s}=13$\,\TeV{} collected during the years 2017 and 2018, which corresponds to more than 70\% of the total data collected by the ATLAS experiment during LHC Run~2.
This limited but significant dataset is used to demonstrate the power of this reconstruction technique; there are no restrictions that would prevent scaling this approach to larger datasets, although it does take a larger amount of analysis-level computing resources than trigger-based analyses.
The data are selected with a single-electron~\cite{TRIG-2018-05} or a single-muon~\cite{TRIG-2018-01} requirement at the HLT, with \pT{} thresholds of 26\,\GeV{} in both cases.
The highest-rate single-electron trigger and single-muon trigger are used, as these two triggers represent a significant fraction of the entire ATLAS trigger rate.
Moreover, single-electron and single-muon triggers are very clear trigger signatures, and it is therefore easy to identify and remove the \tpv{}; the \tpv{}-removal procedure is discussed in detail in \Section{sec:trigger-bias}.

A second dataset is used for validation purposes. It is referred to as the \textit{zero bias dataset}, and it only contains data taken during filled bunch crossings.
It consists of \ppp{} collisions selected exactly one LHC orbit after a reference trigger is fired, thus evolving with the per-bunch luminosity conditions, while remaining independent of the presence or absence of activity in the bunch crossing to be recorded.
This strategy provides a representative sample of \ppp{} collisions unbiased by any of the standard trigger selections~\cite{TRIG-2019-04}.
Data quality requirements are applied to ensure all detector components are operating within acceptable parameters~\cite{DAPR-2018-01} for both the pile-up and zero bias datasets.

Monte Carlo simulation~\cite{SOFT-2010-01} is used to generate samples of events produced by multijet processes. They were generated using \PYTHIA[8.230]~\cite{Sjostrand:2014zea} with leading-order matrix elements for dijet production that were matched to the parton shower. The renormalisation and factorisation scales were set to the geometric mean of the squared transverse masses of the two outgoing particles in the matrix element, \(\hat{p}_\textrm{T} = \sqrt{(\pTX[2][1] + m_1^2) (\pTX[2][2] + m_2^2)}\). The \NNPDF[2.3lo] parton distribution function (PDF) set~\cite{Ball:2012cx} was used in the matrix element generation, the parton shower, and the simulation of the multi-parton interactions. The A14~\cite{ATL-PHYS-PUB-2014-021} set of tuned parameters was used.

In order to account for the effects of multiple \ppp{} interactions, additional interactions were generated using \PYTHIA[8.186]~\cite{Sjostrand:2007gs} with the A3 set of tuned parameters~\cite{ATL-PHYS-PUB-2016-017} and the \NNPDF[2.3lo] PDF set.
A random selection of additional \ppp{} interactions are added to the dijet event described above, with a multiplicity designed to match the distribution of the average number of interactions observed in data.
Additional \ppp{} interactions are added both \textit{in time} to the same simulated bunch crossing, and \textit{out of time} to bunch crossings before or after the one containing the interaction of interest.
All of the resulting particles, both from the process of interest and the additional \ppp{} interactions, are passed through a detailed simulation of the ATLAS detector based on \GEANT{}~\cite{Agostinelli:2002hh}.
The detector contributions of additional \textit{in time} \ppp{} interactions are treated the same as the process of interest, while \textit{out of time} contributions are accounted for according to the nature of each detector.  The ID and MS readout is complete within a single \bc{}, thus \textit{out of time} signals do not appear.  Some of the calorimeters have a long readout time, and thus \textit{out of time} signals are weighted to account for the change in signal response as a function of time: such signals can therefore appear in multiple \bcs{}, but with a reduced energy relative to what they would have had as an \textit{in time} energy deposition.

Splitting the proccess by which simulation is produced into two pieces introduces an artificial difference between data and simulated events.
In data, every \ppp{} collision comes from the same underlying distribution, while Monte Carlo simulated samples use different methods to model the singular collision of interest and the rest of the interactions.
This distinction in samples of simulated events comes from the need to generate individual process(es) of interest, but also to include the impact of additional inclusive collisions on signal reconstruction.
These processes are tuned separately to improve the precision with which they describe data, and thus the same physical process is simulated slightly differently depending on if it is the collision of interest or an additional interaction.
In order to avoid such artificial differences, this paper will only consider the collision of interest when working with simulated events, as is discussed further in \Section{sec:reco:jets}.


%
%
%
\section{Luminosity of the pile-up dataset}
\label{sec:lumi}

%

The integrated luminosity of the standard dataset collected with the ATLAS detector for physics analyses is well defined, and it was measured with very high precision~\cite{DAPR-2021-01}. Differences between the delivered and recorded integrated luminosities are due to bunch crossings where data could not be recorded for different reasons, such as trigger dead time~\cite{TRIG-2019-04} or the detector not being operative.

In the case of the pile-up dataset, there is no trigger evaluating every \ppp{} collision delivered by the LHC: it is only possible to study data that has already been recorded. The total integrated luminosity of this new dataset will therefore be a fraction of the integrated luminosity delivered by the LHC, determined by the proportion of the \rbcs among all delivered bunch crossings. This fraction is equal to the ratio of the trigger rate and the rate of collisions at the LHC.

As discussed in \Section{sec:datasamples}, the pile-up dataset only considers bunch crossings recorded with either a single-electron or single-muon trigger requirement.
The relevant trigger rate, $\rho$, for the dataset is therefore $\rho^{e\cup\mu}$, where $e\cup\mu$ represents the logical-or of the two relevant triggers.
This rate will reduce with time as a consequence of the LHC beams decreasing in intensity as they age and as protons in the beam are used for collisions.
The data-taking period is from the point at which the beams are injected, preceeding first collisions, until the beams are discarded. %
Runs typically last many hours, and are divided into time intervals named luminosity blocks, which have a typical duration of one minute and which are defined as intervals of constant luminosity~\cite{TRIG-2019-04}.

In order to evaluate the rate of the selected triggers, $\rho^{e\cup\mu}$, the number of \rbcs{} in a luminosity block is counted, and this value is divided by the luminosity block duration.
The trigger rate in an example ATLAS run is shown in \Figure{fig:TriggerRatesLumi}, including both the combined rate $\rho^{e\cup\mu}$ and the individual trigger rates.
The total rate $\rho^{e\cup\mu}$ is very slightly smaller than the sum of the individual rates $\rho^{e}+\rho^{\mu}$.
This is expected: some SM processes can naturally produce two different-flavour leptons, such as top quark pair production or diboson production.
Moreover, both triggers and offline reconstruction algorithms have their limitations; it is possible to reconstruct a detector signal as a lepton even in the absence of a real lepton, and thus some different-flavour \rbcs{} may be a mixture of one real lepton and one fake lepton.

The rate of LHC bunch crossings, $\rho^\text{LHC}$, is given by the operational parameters of the accelerator and, while it can change from run to run, it remains constant over the course of a run. It is given by the product of the number of filled \ppp{} bunches for a given run and their rotation frequency of 11245\,Hz~\cite{Gareyte:1995fx}, and typically takes values on the order of 30~MHz, with variations according to the specific LHC filling scheme. The total integrated luminosity for the pile-up dataset can therefore be obtained as
\begin{equation}
\mathcal{L}^\text{PU} = \frac{\rho^{e\cup\mu}}{\rho^\text{LHC}}\times \mathcal{L}^\text{LHC}=\sum_i^{N_\mathrm{runs}} \sum_j^{N^i_\mathrm{LBs}} \frac{\rho^{e\cup\mu}_{i,j}}{\rho^\text{LHC}_{i}}\times\mathcal{L}^\text{LHC}_{i,j},
\label{eqn:lumiPU}
\end{equation}
where $N_\mathrm{runs}$ is the number of runs in the dataset, $N_\mathrm{LBs}^{i}$ is the number of luminosity blocks in run $i$, and $\mathcal{L}^\text{LHC}_{i,j}$ is the LHC delivered integrated luminosity in luminosity block $j$ of run $i$.
This equation considers each individual luminosity block of the run separately, rather than aggregating over the full run, to better account for the time dependence of the trigger rate.
The computed integrated luminosity for each luminosity block for the same example run is shown in \Figure{fig:LumiExampleRun}.
For this run the LHC bunch crossing rate was $\rho^\text{LHC} = 28.6$\,MHz and the total integrated LHC delivered luminosity for this run is 0.42\,\ifb{}, while the total integrated luminosity for the pile-up dataset is 4.0\,\inb{}.

\begin{figure}[t]
\centering
\subfigure[]{\includegraphics[width=0.48\textwidth]{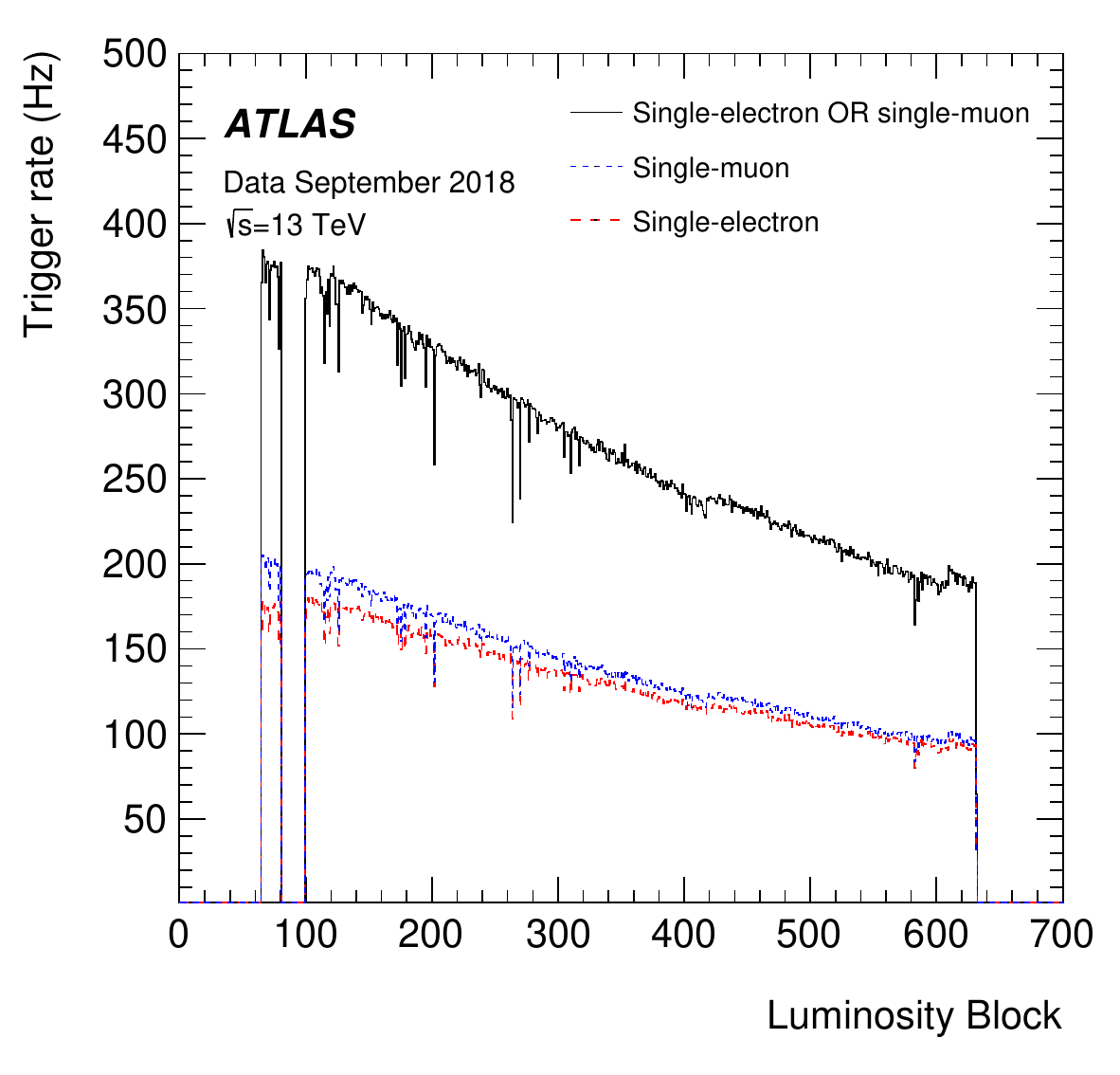}\label{fig:TriggerRatesLumi}}
\subfigure[]{\includegraphics[width=0.48\textwidth]{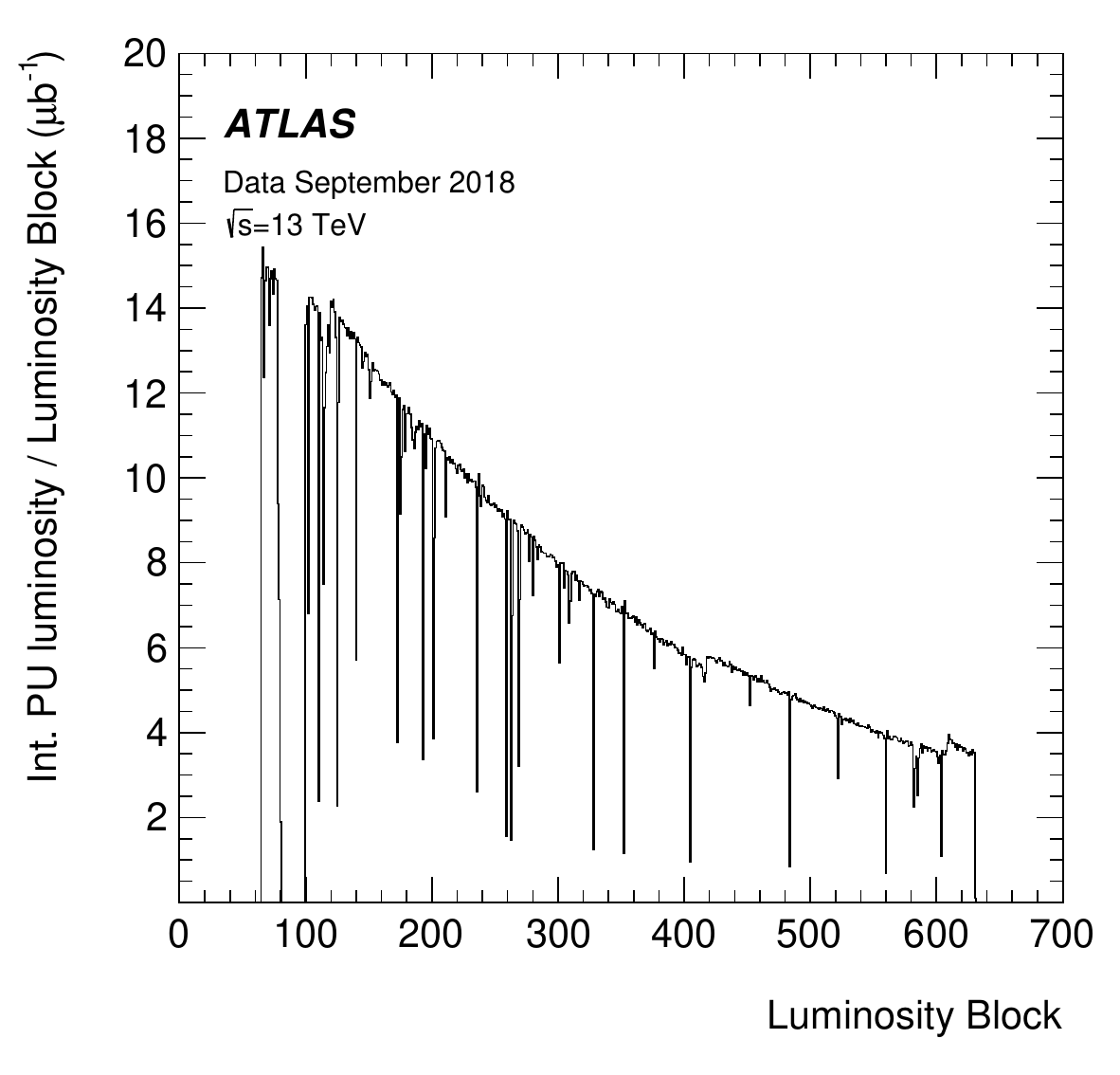}\label{fig:LumiExampleRun}}
\caption{The integrated luminosity calculation for the pile-up dataset in an ATLAS run taken in September 2018, with the (a) rate of the single-electron and single-muon selections at the trigger level, individually (dashed lines) and combined (solid line), and (b) computed integrated luminosity per luminosity block. Both plots are shown as a function of the luminosity block. The total integrated luminosity for the pile-up dataset in this run is 4.0~nb$^{-1}$. Short-lived reductions in the rate and integrated luminosity arrise from detector operational issues.
}
\end{figure}

After applying this procedure to the full dataset, the pile-up integrated luminosity is calculated to be 0.47\,\ipb{} in 2017 and 0.59\,\ipb{} in 2018, compared with 44.6\,\ifb{} and 58.8\,\ifb{} respectively for the triggered dataset.
The full 2015--2018 dataset is determined to have an integrated luminosity of 1.33\,\ipb{} for pile-up, compared with 140\,\ifb{}.
As the procedure used to scale the integrated luminosity is based on the actual trigger rate in the recorded data, the calculation is deterministic, and does not introduce additional uncertainty components.
The integrated luminosity uncertainty for the pile-up dataset is therefore the same as that of the triggered dataset, thus the uncertainties are 1.13\% for 2017, 1.10\% for 2018, and 0.83\% for 2015--2018~\cite{DAPR-2021-01}.

The impact of not using every \ppp{} collision in the \rbc{}, such as the fact that the \tpv{} is always discarded, results in a reduction of the number of events available for physics analysis. On the other hand, the possibility of having more than one useful \pv{} in a given \rbc{} has the opposite effect. Both of these effects are considered as selections applied on the \rbcs{} and do not impact the initial size of the dataset to analyse, and therefore are not considered as part of the integrated luminosity calculation. Instead, they are treated as part of the efficiency of the dataset reconstruction and are discussed in \Section{sec:PV-selections}.

With this first measure of the pile-up integrated luminosity, it is possible to evaluate the potential of using the pile-up dataset rather than conventional trigger-based datasets.
A series of single-jet triggers, with transverse energy $E_T$ thresholds as low as 15\,\GeV{} and as high as 420\,\GeV{}, are used to cover the full kinematic range to which the ATLAS detector is sensitive.
The highest-threshold trigger is unprescaled: it sees and can trigger on all of the LHC-delivered \ppp{} collisions, thus it corresponds to the full integrated luminosity of 140\,\ifb{} as recorded by ATLAS during Run~2.
It is not possible to record all of the collisions that are selected by the triggers of lower-threshold single-jet triggers as the inclusive jet production cross-section is too large, and it would overwhelm the ATLAS data acquisition system.
Instead, all but the highest-threshold trigger are prescaled: they are only active for a fraction of the delivered bunch crossings, and thus have a correspondingly reduced integrated luminosity.
Each prescaled trigger is associated with a prescale factor $N$ such that on average only $1/N$ of the bunch crossings that would have been selected by the trigger is recorded.
The value of $N$ for each of the single-jet triggers evolves during the run to follow the evolution of the instantaneous luminosity, therefore allowing for the triggers to record bunch crossings at a given rate across the full run, typically at the level of a few Hz.

The single-jet triggers with the lowest $E_T$ thresholds have a corresponding effective integrated luminosity significantly lower than that of the pile-up dataset, as shown in \Figure{fig:lumi_pile-upVsTrig}.
Neglecting the efficiency of using each dataset for a given analysis, the integrated luminosity comparison provides a rough regime of relevance for the pile-up dataset: the pile-up dataset should have superior statistical sensitivity whenever the analysis in question relies on events where the most energetic jet (the leading jet) has a \pT{} below roughly 70\,\GeV{} (60\,\GeV{}) in 2018 (2015--2018).
The regime of relevance grows when considering only 2018, as the 2018 trigger rate averaged 1.2\,kHz instead of 1.0\,kHz in 2015--2017, and 2018 also had a higher average number of inelastic \ppp{} collisions per \bc{} than previous years.

\begin{figure}
\centering
\subfigure[]{\includegraphics[width=0.48\textwidth]{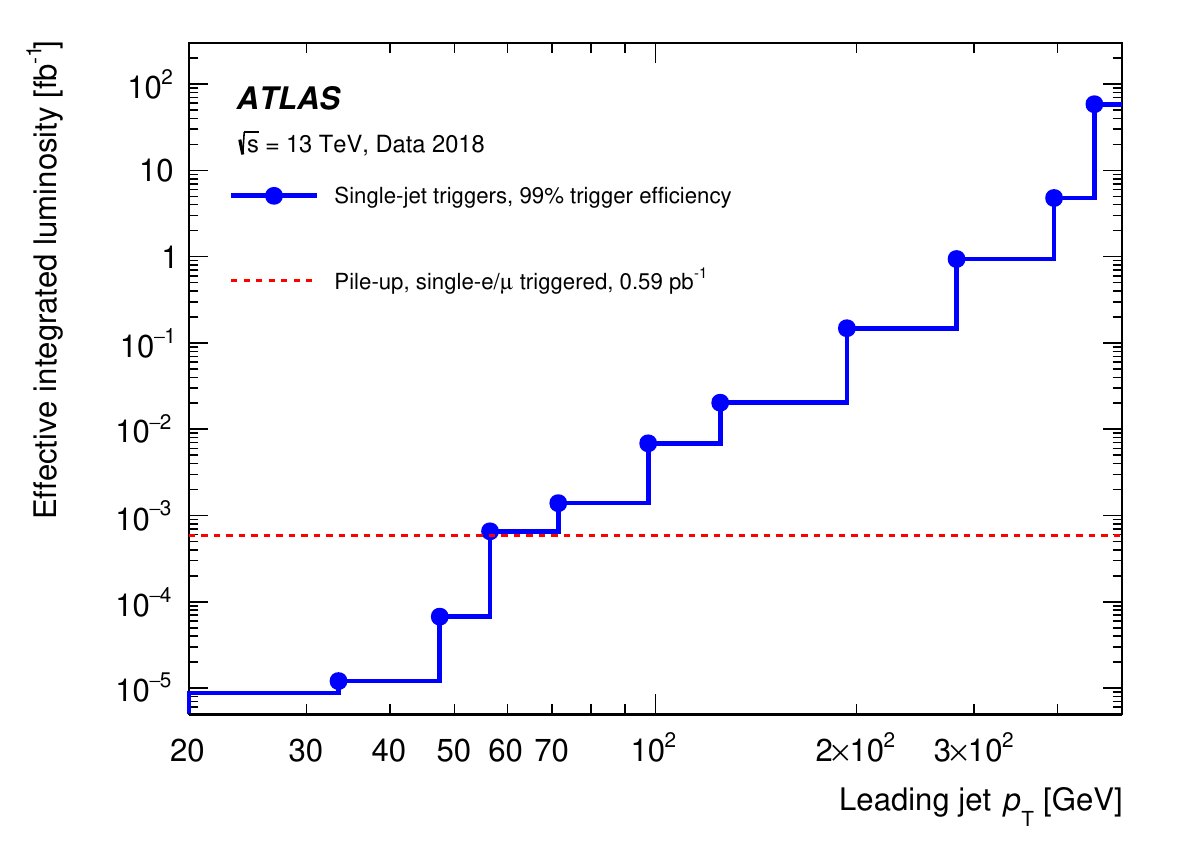}\label{fig:lumi_pile-upVsTrig_2018}}
\subfigure[]{\includegraphics[width=0.48\textwidth]{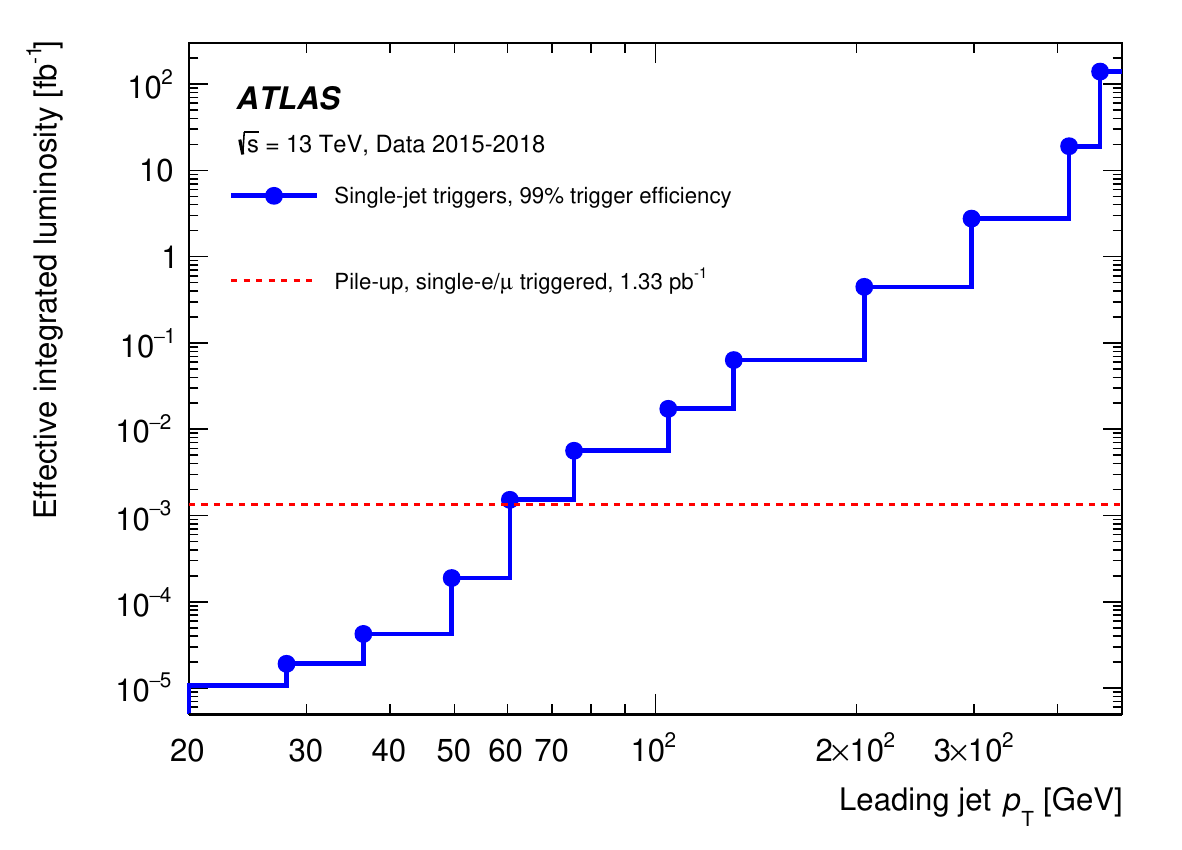}\label{fig:lumi_pile-upVsTrig_Run2}}
\caption{
The integrated luminosity recorded by ATLAS using a series of single-jet triggers in (a) 2018 and (b) 2015--2018, shown for the point at which the trigger is 99\% efficient as a filled circle.
The point at which the trigger is 99\% efficient denotes the start of that dataset, thus there is a fixed integrated luminosity until the start of the next trigger; this is represented as a solid line showing the kinematic regime covered by the trigger in question.
The single-jet-triggered dataset is compared with the integrated luminosity of the pile-up dataset, showing the integrated luminosity for the choice of single-electron and single-muon triggers taken in this result as a horizontal dashed line.
}
\label{fig:lumi_pile-upVsTrig}
\end{figure}

As mentioned in \Section{sec:datasamples}, ATLAS also records a sample of zero bias data, which could also be used for the same purpose as the pile-up dataset.
This dataset, however, is much smaller: in 2018, the zero bias trigger had a typical rate slightly below 10\,Hz, and the total integrated luminosity of the 2018 zero bias dataset amounts to 19.3\,\inb{}.
The 2018 pile-up dataset is therefore roughly 30 times larger than the zero-bias-triggered dataset, further motivating the relevance of the pile-up-based approach.
Roughly 300\,Hz of zero-bias data would be necessary to match the pile-up dataset, and while ATLAS could record such a large amount of zero-bias data, it is not worth the cost to the larger ATLAS physics programme.
The pile-up dataset is thus an excellent means of obtaining a large trigger-unbiased dataset without impacting the trigger rate or storage costs, thus avoiding any negative impacts on other physics objectives.


%
\section{Reconstructing individual collisions}
\label{sec:reco}

%

The reconstruction of the pile-up dataset requires the ability to differentiate between separate \ppp{} collisions in the same bunch crossing, and to match individual detector signals to the originating collision.
This is only possible through the usage of charged-particle tracks reconstructed in the ID, as the calorimeter does not have the necessary spatial resolution to differentiate between separate collisions.
These charged-particle tracks are used to form \pvs{}, and each \pv{} is assumed to correspond to a single \ppp{} collision, as vertex merging is rare, reaching the level of 2.6\% in high-pile-up conditions in June 2018~\cite{IDTR-2018-006}.
The impact of vertex merging on the results in this paper are further suppressed by dijet analysis selections, as described in \Section{sec:eventint}, and by the even tighter dijet requirements in \Section{sec:JER_details}.

Tracks from individual charged particles, such as electrons and muons, can therefore be linked to their originating \pv{} with near-perfect accuracy.
Photons, as purely neutral physics objects, generally do not appear in the ID and therefore cannot be matched to a given \pv{}; they therefore have limited use in the pile-up dataset.
Some photons, however, convert into electron-positron pairs; depending on when the conversion occurs, these photons may be possible to match to a given \pv{}.
Jets, as a composite object comprised of many charged and neutral particles, can be matched to a specific \pv{} following a more advanced treatment.

Due to the orders-of-magnitude difference between SM cross-sections for charged lepton production and inclusive jet production, the use of these types of physics objects differs dramatically within the context of the pile-up dataset.
The cross-section for the production of prompt charged leptons is rare, and thus preferentially selected by the ATLAS trigger strategy; charged leptons are therefore primarily used to identify the \tpv{}(s), which are intended to be rejected when building the pile-up dataset.
In contrast, hadronic physics processes are the dominant source of pile-up collisions; jets are therefore of primary interest when working with \ppvs{}, which are intended to be used for subsequent data analysis.
Therefore while all physics objects are important for this analysis, their intended use differs substantially. %

\subsection{Standard object reconstruction}
\label{sec:reco:standard}

The trajectories of charged particles in the ID are reconstructed as tracks by fitting a trajectory model to the hits measured in the tracker~\cite{PERF-2015-08}.
Reconstructed tracks with $\pT^{\text{track}}>400$\,\MeV{} are used as input for primary vertex reconstruction after a set of quality requirements.
An iterative process first finds a seed for a vertex, the position of which is then refined by iteratively minimizing the distance between the vertex and the tracks pointing at it~\cite{PERF-2015-01}.
Once a vertex is reconstructed, its associated tracks are removed from the track pool used as input to the algorithm and the vertex reconstruction process is repeated to find more \pvs until no more tracks are left or no more vertices can be found. The \pvs considered in the jet reconstruction procedure are required to have at least two associated tracks with $\pT^{\text{track}}>500$\,\MeV{}.
Vertices are sorted from highest to lowest \pTvtx{}, thus the vertex with the highest \pTvtx{} is referred to as \pv{}0, the second highest as \pv{}1, and so on.

Particle showers in the calorimeters are reconstructed with the topological clustering algorithm into \textit{topo-clusters}~\cite{PERF-2014-07}.
They are seeded by calorimeter cells for which the absolute energy measurement is at least four times the expected noise.
The topo-cluster is expanded by adding all adjacent cells, and this expansion iteratively repeats for the neighbours of all newly added cells with an energy measurement at least two times the expected noise, at which point a last layer of all adjacent cells is included.

Particle flow objects (PFOs) are reconstructed combining ID tracks and topo-clusters. A detailed description of the procedure can be found in Ref.~\cite{PERF-2015-09}.
The particle flow algorithm starts with tracks, representing charged particles, and attempts to match each track to a single topo-cluster; this is done iteratively, from high- to low-momentum tracks.
The energy expected to be deposited in the calorimeter by the charged particle is estimated to decide whether additional topo-clusters should be considered.
This estimated energy is then subtracted cell-by-cell from the matched topo-cluster(s); the track is retained in all cases, while energy-subtracted topo-cluster(s) are only retained if they still have a significant energy after the subtraction procedure.
The tracks are then classified as \textit{charged PFOs}, while the remaining topo-clusters not matched to a track, or that remain after the energy-subtraction procedure, are classified as \textit{neutral PFOs}.

Electrons are reconstructed by combining information from the electromagnetic calorimeter and the ID~\cite{EGAM-2018-01}. They are required to have $\pT{}>7\,\GeV{}$ and $|\eta|<2.5$, and to satisfy the tight likelihood identification point together with tight isolation requirements~\cite{ATLAS:2023dxj}.
Photons are reconstructed using clusters of energy deposits in the electromagnetic calorimeter~\cite{EGAM-2018-01}. Photons are required to have $\pT{}>7\,\GeV{}$ and $|\eta|<2.37$, and are not considered if they fall within the transition region between the barrel and endcap calorimeters ($1.37 < |\eta| < 1.52$).  They are required to satisfy the tight likelihood identification point with tight isolation requirements~\cite{ATLAS:2023dxj}.
Muons are reconstructed combining measurements from the ID and the MS~\cite{MUON-2018-03}. Different techniques are used to reconstruct muon candidates, depending on the number of hits in the ID and MS stations and their $\eta$ coordinate in the detector. The reconstructed muons are required to have $\pT{}>7\,\GeV{}$ and $|\eta|<2.5$ and to satisfy the tight identification point and tight track-based isolation criteria~\cite{MUON-2018-03}.

\subsection{Jet reconstruction in the pile-up dataset}
\label{sec:reco:jets}

Jets in the pile-up dataset are clustered following the same approach as a single-\pv{} analysis in ATLAS: using the \antikt{} algorithm~\cite{Cacciari:2008gp}, with a radius parameter $R=0.4$, as implemented in the \textsc{FastJet}~\cite{Fastjet} software package.
The difference is instead in how the inputs to \textsc{FastJet} are prepared, the use of \textsc{FastJet} to cluster jets once per \pv{} rather than once per \rbc{}, and how the subsequent jet attributes are calculated.

In single-\pv{} ATLAS jet reconstruction, PFOs are used as inputs, and a single \pv{} of interest is identified; \pv{}0 is almost always the \pv{} of interest.\footnote{There are rare exceptions to this, such as di-photon analyses as in Ref.~\cite{HIGG-2019-16} or analyses where the visible activity is outside of the tracker. In such cases, an alternative \pv{} can be used, but jet reconstruction is still performed for a single \pv{} per bunch crossing.}
All of the neutral PFOs are considered, as they are all candidates to have originated from the \pv{} of interest, and their four-vectors are corrected to be consistent with originating from that \pv{}.
The charged PFOs do have a clear \pv{} interpretation, and thus only charged PFOs satisfying $|\Delta z(\mathrm{track},\mathrm{vertex})\sin\theta| < 1\,\mathrm{mm}$ are used, where $\Delta z$ is the distance of closest approach of the track to the \pv{} along the beam axis.
This selection retains tracks from the collision of interest, while removing undesired charged hadronic activity from other \pvs{}~\cite{PERF-2015-09}.
This set of origin-corrected neutral PFOs, and \pv{}-of-interest-consistent charged PFOs, is then grouped into a single set of four vectors, which are used as inputs to \textsc{FastJet}.

In order to reconstruct jets from pile-up collisions, the single-\pv{} approach was extended to be performed with respect to all \pvs{}, rather than a single \pv{} of interest, following a similar approach to Ref.~\cite{ATL-PHYS-PUB-2019-026}.
The neutral PFOs still have no natural \pv{} interpretation, and thus are duplicated by the number of \pvs{}, where each copy is origin-corrected to point to a different \pv{} and marked as being linked to that \pv{}.
The same charged PFO selection is appropriate, when performed on a vertex-by-vertex level, with one change: if there are multiple vertices satisfying the selection, the charged PFO is only used for the vertex that minimises $|\Delta z(\mathrm{track},\mathrm{vertex})\sin\theta|$, thus avoiding double-usage of any charged PFO\@.
Each \pv{} is then processed sequentially, where the corresponding origin-corrected neutral PFOs and selected charged PFOs are used as inputs to \textsc{FastJet} for jet clustering, producing a distinct set of jets built with respect to each \pv{}.
As a summary, the following steps are performed during multi-\pv{} jet reconstruction:
\begin{enumerate}
\item Select the current \pv{} with respect to which jets will be reconstructed;
\item Identify all charged PFOs consistent with the current \pv{}, meaning that the current \pv{} minimises $|\Delta z(\mathrm{track},\mathrm{vertex})\sin\theta|$, and that minimised value satisfies a requirement of being less than 1\,mm;
\item Adjust all neutral PFO 4-vectors to point to the current \pv{};
\item Run \textsc{FastJet} clustering for the current \pv{} on this set of charged and neutral PFO inputs;
\item Repeat the procedure for every \pv{}, where the ordering in which the \pvs{} are processed has no impact: each charged PFO is matched to at most one \pv{}, and all neutral PFOs are used for every \pv{}.
\end{enumerate}

In addition to changing the inputs to jet clustering, the subsequent calculations of jet properties have also been adapted.
There are many jet properties that depend on \pv{}-specific quantities, such as the number of tracks associated with a \pv{} that are within the ghost-association-defined catchment area of the jet~\cite{Cacciari:2007fd}: the tracks are included in jet reconstruction as four-vectors of infinitesimal magnitude, and then assigned to the jet with which they are clustered.
These quantities are used for a variety of purposes, such as pile-up mitigation, jet calibration, jet tagging, and more.
All such jet property calculations were updated from being calculated for the \pv{} of interest, to instead being calculated for the \pv{} with respect to which the jet was built.

Jets from pile-up collisions are calibrated using the same procedure as typically used by ATLAS: a series of Monte-Carlo-based calibrations to correct on average to the particle-level scale, followed by \textit{insitu} corrections accounting for differences between data and simulated events~\cite{JETM-2018-05}.
The response of the detector to an incident hadronic shower forming a jet is independent of the presence of other non-overlapping objects in the event, and is thus also independent of the vertex from which the jet originates.
The response does depend on the location of the jet within the detector, the energy scale of the jet, and other properties such as the fraction of charged-particle momentum in the jet, but such dependencies are present independent of the originating \pv{}, and thus the jet calibration is already parameterised with respect to such variables.
There is one possible exception to the claim that jets are independent of the \pv{} they are reconstructed with respect to, which occurs when two or more \pvs{} are so close that they could each satisfy the $|\Delta z(\mathrm{track},\mathrm{vertex})\sin\theta| < 1\,\mathrm{mm}$ requirement.
In this case the single-\pv{} approach will use all of the charged PFOs satisfying the requirement as inputs to jet reconstruction, while all-\pv{} reconstruction will only assign the charged PFO to the closest \pv{}, and thus the reconstructed jet can be slightly different; this possibility was studied and is found to be both extremely rare and to have only a small impact on the reconstructed jet both before and after calibration.

\subsection{Jet selections in the pile-up dataset}
\label{sec:reco:select}

The jet collection associated with each \pv{} in each \rbc{} is filtered to retain only well-understood jets.
Unless otherwise stated, a requirement of $\pT{}>20\,\GeV{}$ is applied to all jets in order to be compatibile with the standard ATLAS jet calibration recommendations.
Additionally, all jets are required to be within the tracker acceptance, $|\eta|<2.4$, to ensure that the charged PFOs used in their reconstructions can be uniquely associated with a \pv.

Jets are then evaluated for their consistency with respect to the interpretation of originating from the \pv{} with respect to which they were reconstructed.
The \textit{jet vertex tagger} (JVT)~\cite{PERF-2014-03} is used for this purpose, where a JVT score is calculated with respect to the current \pv{} of interest.
JVT provides substantial discrimination between jets originating from other \pvs{} or from stochastic overlaps of detector signals, as opposed to jets consistent with originating from the \pv{} of interest.
This is done by identifying all of the tracks ghost-associated with the jet, and evaluating the amount of charged-particle momentum carried by those tracks matched to the \pv{} of interest as opposed to other \pvs{}.
The JVT discriminant additionally compares the charged-particle momentum carried by tracks matched to the \pv{} of interest with the total momentum of the jet in order to refine its classification of the jet as being either consistent or inconsistent with originating from the \pv{} of interest.

The use of JVT is critical to the reconstruction of jets in the pile-up dataset.
Without JVT, most detector signals would be reconstructed numerous times, once with respect to each \pv{}, due to how the neutral PFOs are used in the reconstruction procedure.
A well-isolated detector signal will therefore be reconstructed once with its correct energy, including both charged and neutral PFOs, and $N_\mathrm{PV}-1$ times with a reduced energy, where only the neutral PFOs are included.
The use of JVT removes these duplicates, as only jets with a charged component consistent with the \pv{} of interest are retained.
An example of what may remain after reconstructing each \pv{} in a given \rbc{} is provided in \Figure{fig:jetReco}, showing only the jets that survive the JVT requirement.
To first order, the remaining jets are what would be used in a physics analysis, but there are uncommon cases where two signals with sizeable charge fractions overlap and thus both survive the JVT requirement; the identification and removal of such cases are addressed in \Section{sec:jet-jet-overlaps}.

\begin{figure}
\centering
\includegraphics[width=\textwidth]{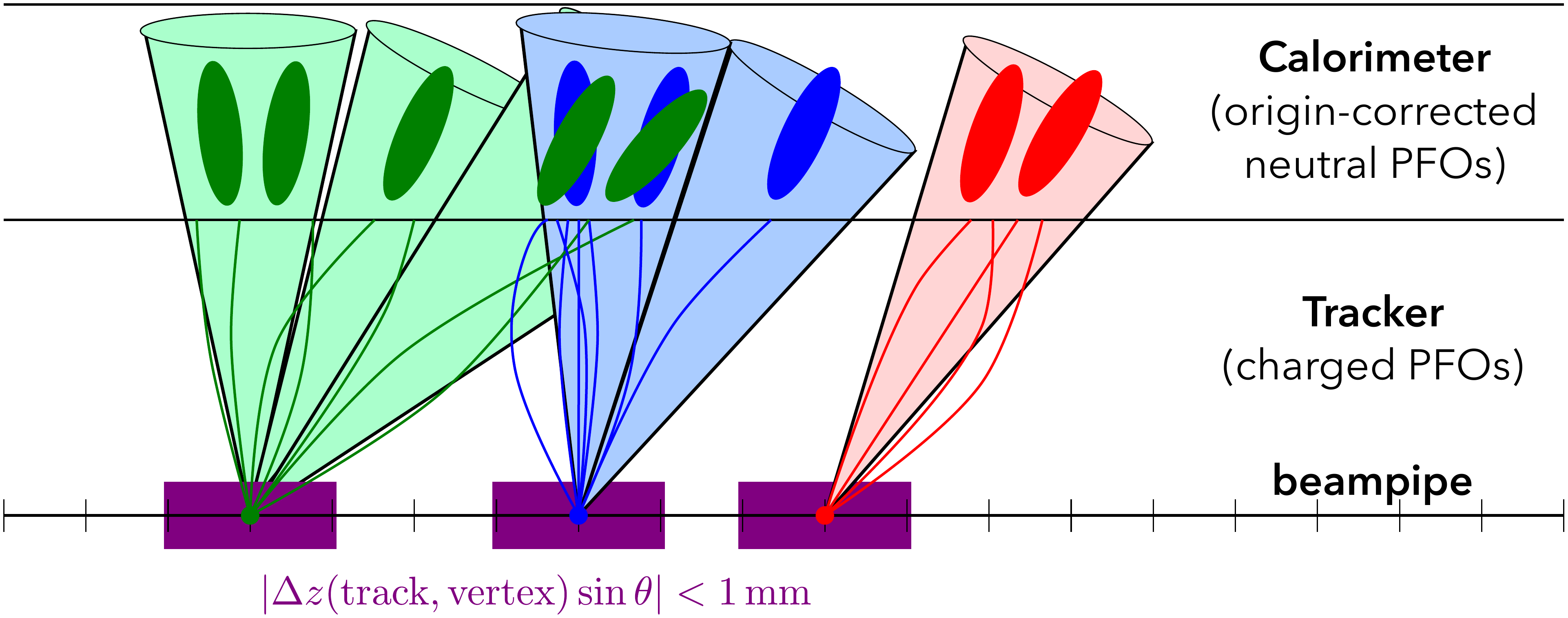}
\caption{
A graphical representation of multi-\pv{} jet reconstruction, as implemented for the pile-up dataset.
In reconstructing a given \pv{}, jets are built from the set of charged PFOs compatible with that \pv{} and the set of all neutral PFOs origin-corrected with respect to the same \pv{}.
The resulting jets will include many neutral-only duplicates, where the charged component indicates that the jet is incompatible with the \pv{}.
The surviving jets are, to first order, well-defined representations of the signals present in the detector.
An exception to this is shown as a pair of overlapping jets, one green and one blue.
These jets correspond to a pair of separate collisions producing independent signals, which happen to overlap in the detector, and thus where both jets satisfy the JVT requirement despite sharing the same neutral PFOs.
}
\label{fig:jetReco}
\end{figure}

\subsection{Jet calibration in the pile-up dataset}
\label{sec:reco:calib}

The applicability of the jet calibration to jets reconstructed with respect to different \pvs{} can be studied in simulated events by comparing the jet \pT{} response, defined as $\mathcal{R}=\pT^\mathrm{reco}/\pT^\mathrm{true}$, for jets from events where the generated process is reconstructed as either the leading or sub-leading vertex in the event (correspondingly \pv{}0 or \pv{}1).
The reconstructed jets ($\pT^\mathrm{reco}$) are geometrically matched to particle jets ($\pT^\mathrm{true}$), where the particle jets are built using stable ($c\tau>10\,\mathrm{mm}$) final-state particles from the generated process, excluding muons and neutrinos.
To be geometrically matched, the distance between the particle and reconstructed jet is required to satisfy ${\Delta}R \leq 0.3$, with an isolation requirement: reconstructed (particle) jets are required to have no other reconstructed (particle) jet of $\pT{} > 7\,\GeV{}$ within ${\Delta}R \leq 0.6$, following the procedure and particle-jet definition as in Ref.~\cite{JETM-2018-05}.

A direct comparison of the \pT{} response of jets from different vertices is not sufficient, as it imposes an implicit bias.
Vertices are sorted by \pTvtx{}, which in turn is correlated with the amount of charged activity in the jet, as is discussed more in \Section{sec:other-BC-activity}.
The amount of charged activity in the jet is related to the jet \pT{} response, and thus the \pTvtx{} ordering impacts the comparison of jets from different vertices.
By imposing a selection on \pTvtx{}, one can remove this implicit bias and focus on whether there is a dependence originating from the presence of other higher-energy activity in the bunch crossing.
The exact value of the selection is not important, rather it is important to impose a requirement of a similar \pTvtx{} on the vertices being compared.
\Figure{fig:jetResponse} demonstrates that there is no substantial response dependence on the \pv{} the jet originates from, which supports the statement that the \pv{}0-derived jet calibration is appropriate for jets from \ppvs{}.

\begin{figure}
\centering
\includegraphics[width=0.48\textwidth]{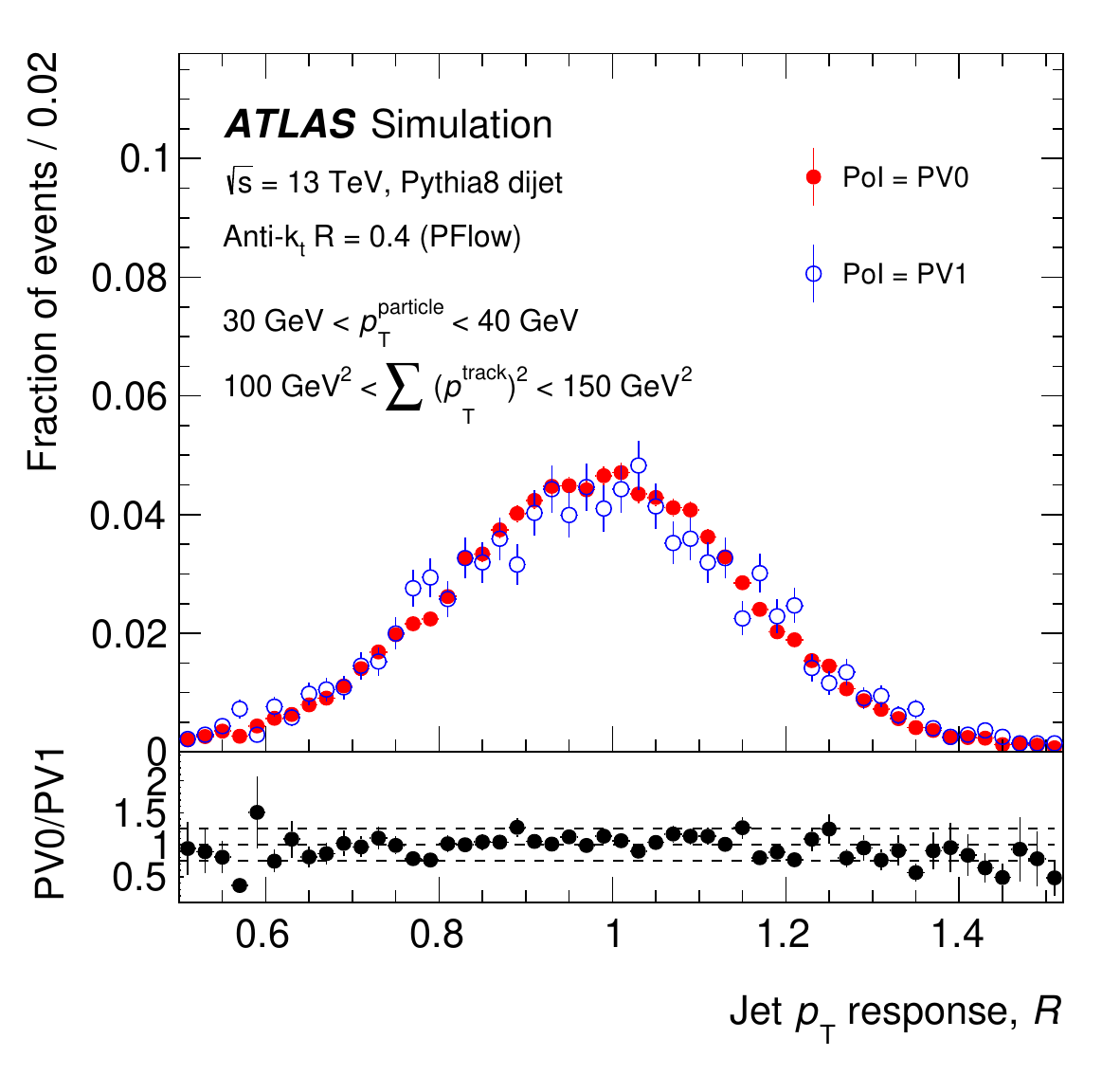}
\caption{
The dependence of the isolated and particle-jet-matched \pT{} response, $\mathcal{R}$, on whether the generated process of interest (\poi{}) is the most energetic (\pv{}0) or second most energetic (\pv{}1) reconstructed collision in the bunch crossing.
A selection is applied on \pTvtx{} to mitigate vertex ordering effects, thus focusing on whether the presence of other higher-energy processes in the same bunch crossing impacts the detector response to jets.
}
\label{fig:jetResponse}
\end{figure}


%
\section{Reconstructing bunch crossings}
\label{sec:eventint}

%
The different \ppp{} collisions taking place within a given bunch crossing are all physically independent processes.
However, their interaction with the detector and their consecutive reconstruction can lead to correlated phenomena across different \pvs.
In order to create a dataset that is suitable for physics analysis use, the different sources of possible crosstalk among reconstructed collisions need to be understood and mitigated as much as possible.

Each of the simultaneous \ppp{} collisions in a given \rbc{} will independently produce a series of particles.
This may include both collimated particle showers originating from a hard-scatter processes, with a momentum scale that is large compared to the proton mass, and additional lower-energy interactions between other incident partons in the same \ppp{} collision, typically referred to as the underlying event~\cite{Butterworth:2012fj}.
This is repeated for each collision, leading to a large number of particles incident on the detector.
Hard-scatter processes are typically distributed in localised regions, and if they have sufficient energy they are reconstructed as independent physics objects, such as jets.
Particles from the underlying event are more diffusely distributed, and are also predominantly at an energy scale below the threshold at which individual physics objects are reconstructed.
The sum over the underlying event from all of the \ppp{} collisions, together with very low-energy hard-scatter processes, therefore represents a stochastic detector background; this background contributes a relatively uniform energy density across the calorimeter, especially in the central region used for pile-up-based jet reconstruction.
The overlap of a jet with this ambient background is traditionally treated as a jet calibration task; an event-by-event and jet-by-jet pile-up subtraction based on jet areas is performed, followed by average pile-up expectation corrections~\cite{JETM-2018-05}.
As this calibration was derived under a \pv{}0-based reconstruction strategy, it is important to cross-check that it also applies to other \pvs{}; this was done in \Section{sec:reco:calib}.

The overlap of separate high-momentum signals from different collisions, each of which has sufficient momentum to create a jet, is not typically considered in single-\pv{} analyses.
Such situations are an important possibility to be mitigated in the pile-up-based approach, and are discussed in \Section{sec:jet-jet-overlaps}.
Following a careful treatment of overlapping high-momentum signals, each jet in the detector should be independent.
The next step is to identify and remove the triggering collision(s) from the \rbc{}, thus removing any bias on the \ppvs{} related to the trigger decision; \Section{sec:trigger-bias} addresses this task.
A further set of selection criteria are then applied to the \ppvs{} to ensure they correspond to well-understood and independent hard-scatter processes, as described in \Section{sec:PV-selections}.
The properties of jets from the the selected \ppvs{} are then compared in \Section{sec:other-BC-activity} in order to ensure that there are no remaining biases, and example \rbcs{} are visualised in \Section{sec:reco:BC}.

\subsection{Mitigating overlapping high-momentum signals}
\label{sec:jet-jet-overlaps}

In the case of two overlapping high-momentum processes from independent hard-scatter interactions, originating from separate collisions and thus separate \ppvs{}, the charged PFOs from each are uniquely associated to their origin \ppv{} and therefore only considered in the corresponding jet reconstruction process, leading to two distinct jets.
The neutral PFOs will, conversely, be pooled together without the possibility of identifying the \ppvs{} from which they originated.
As such, they are included in the reconstructed jets from both of the contributing \ppvs{}, resulting in jets with double-counted neutral PFO energy.
Assuming that both of the jets from the two \ppvs{} are of a comparable momentum scale, each will still have a considerable amount of charged energy, thus surviving the JVT requirement.
The same reasoning extends to cases where three or more comparable high-momentum signals overlap in the detector.

In order to understand the frequency and impact of these overlaps, a new property is defined for each jet as the ratio of the jet \pT ($\pT^{\text{probe}}$) as reconstructed with respect to the current \pv{} (CPV), and the scalar sum of the \pT of all selected jets reconstructed with respect to alternative \pvs{}, within a $\Delta R^\prime = \sqrt{(\Delta\eta)^2+(\Delta\phi)^2}<0.4$ cone from the axis of the probe jet ($\sum{\pT^{\text{overlap}}}$),
\begin{equation}
f_{\mathrm{jet}}^{\mathrm{CPV}} = \frac{\pT^{\text{probe}}}{\pT^{\text{probe}} + \sum{\pT^{\text{overlap}}}}.
\end{equation}
This quantity is sensitive to the momentum scale of the jets that are selected: as the \pT{} threshold is lowered, the calorimeter is increasingly populated with low \pT{} jets, and thus the overlap frequency increases.
While the normal jet \pT{} threshold considered for the pile-up dataset is 20\,\GeV{}, in this case the jet \pT{} threshold was lowered to 15\,\GeV{} in order to reduce threshold effects; this choice is more conservative, as $\pT{}\geq20\,\GeV{}$ jets are now marked as overlapping in the presence of lower-energy-scale surroundings.
For jets reconstructed from a unique high-momentum signal in the detector, the value of \fCPV is exactly 1, while for jets originating from overlapping scenarios, their \fCPV{} value is smaller than 1.
As shown in \Figure{fig:ProbOverlaps}, the fraction of jets originating from the overlap of multiple high-momentum signals decreases with $\pT^{\text{probe}}$.
This is expected, since the probability of having two high-\pT overlapping jets is much smaller than two low-\pT overlapping jets.
It is likely that low-\pT signals could overlap with a high-\pT one, but in these cases the low-\pT{} signal would be reconstructed as a high-\pT jet with very low charged fraction, and therefore would be rejected in the jet-vertex association step discussed in \Section{sec:reco:BC}.

The \fCPV{} distribution shown in \Figure{fig:fCPVDistr} exhibits a large peak at 0.5 due to the fact that overlapping signals in the detector cannot be disentangled.
Due to neutral PFOs being shared in the jet reconstruction algorithm, when reconstructing both signals with respect to their origin \pv{}, the total energy seen by both is roughly the same and is equivalent to the sum of the energy deposited by both signals.
Two jets from different \pvs{} are therefore reconstructed with similar energies, and at the same location in the detector, so the value of \fCPV{} for each of them would be equal to their individual transverse momentum divided by twice that amount, or approximately 0.5.
The additional selection on JVT further enhances this feature in the \fCPV{} distribution by removing low-momentum jets in cases where they overlap with a shower with much larger energy. A second peak can be seen in the distribution at around 0.3, hinting at cases where three similarly energetic signals overlap in the detector.

\begin{figure}[t]
\centering
\subfigure[]{\includegraphics[width=0.48\textwidth]{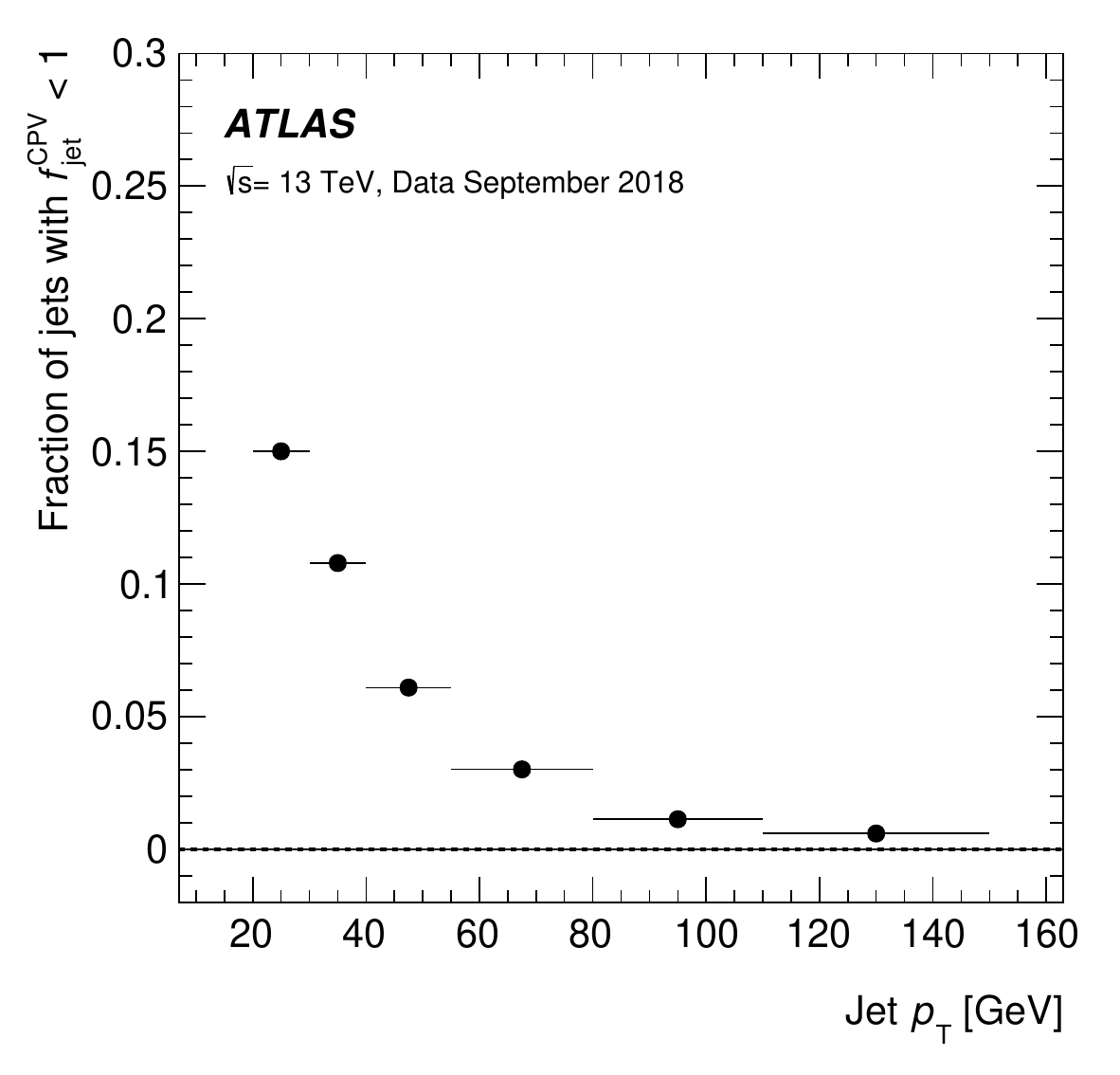}\label{fig:ProbOverlaps}}
\subfigure[]{\includegraphics[width=0.48\textwidth]{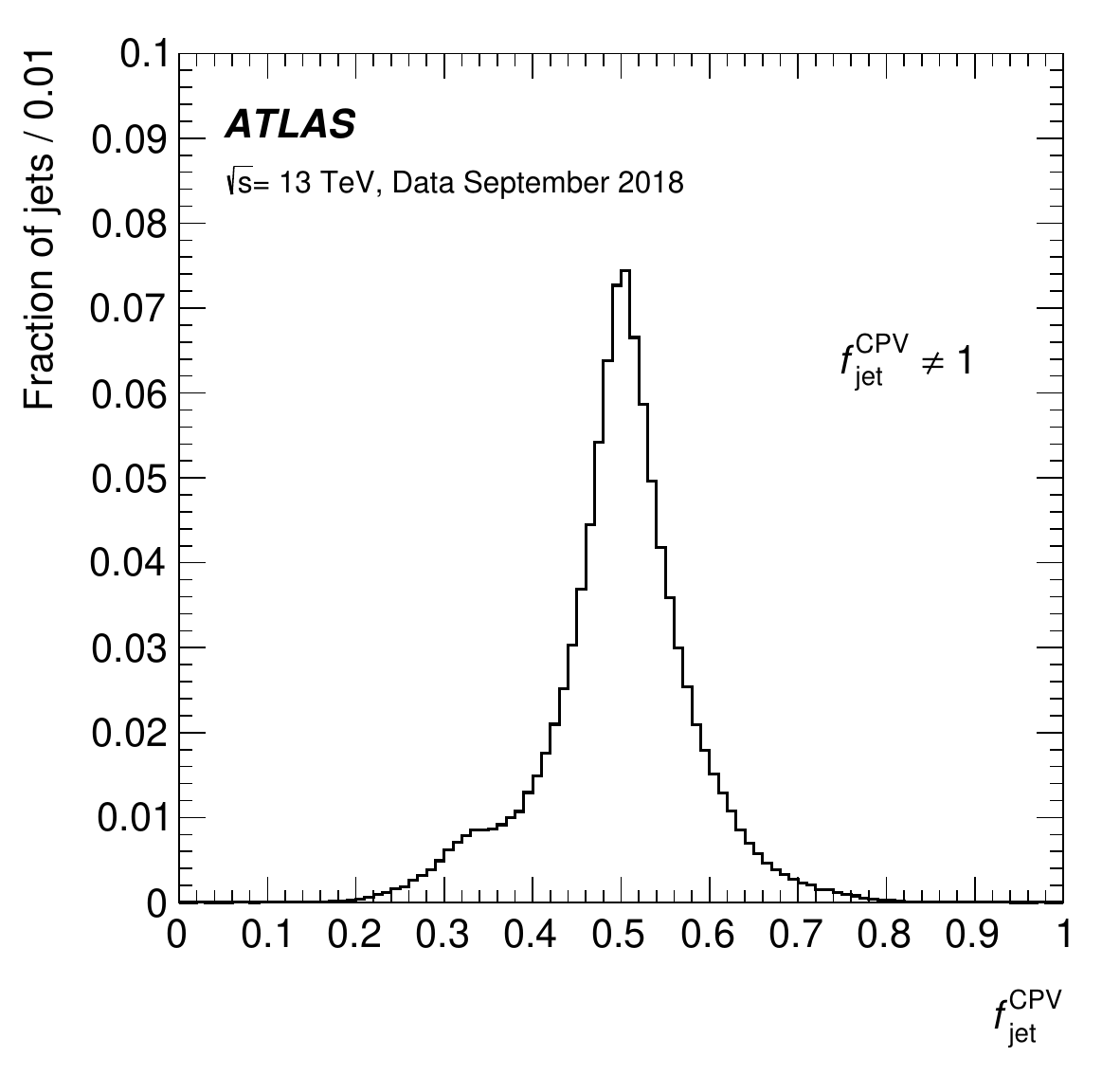}\label{fig:fCPVDistr}}
\caption{(a) Fraction of jets reconstructed from overlapping high-momentum signals in the detector, and (b) distribution of the jet overlap \pT fraction, \fCPV, after removing the peak at $\fCPV{}=1$ containing 88\% of selected jets.
The pile-up dataset from a single ATLAS run taken in September 2018 is used, while a minimal selection of at least two jets with $\pT>20$\,\GeV{} and satisfying the JVT requirement is applied. Statistical uncertainties are included, and are negligible.
}
\end{figure}

These studies show that it is uncommon for jets originating from different vertices, which satisfy the JVT requirement, to overlap in the detector ($\fCPV{}\not=1$).
All further studies in this document therefore impose a requirement of $\fCPV{}=1$ on all of the jets that enter the plot.  In some cases, the criterion is even more strict, as will be discussed where appropriate.

\subsection{Independence from trigger decisions}
\label{sec:trigger-bias}

The standard ATLAS approach to recording data, by means of a kinematic or topological selection at the trigger level, translates into a certain bias applied to the \rbcs.
Accessing a trigger-unbiased dataset by reconstruction of the \ppvs{} necessarily requires the removal of any impact from trigger selections.
The fulfilment of this requirement depends on the signature the \rbc{} was triggered by.
Some of the trigger selections used during Run~2 involve global quantities, such as the total energy deposited in the calorimeter, or the overall transverse momentum imbalance (missing transverse momentum) from the collisions~\cite{ATL-DAQ-PUB-2018-002,ATL-DAQ-PUB-2019-001}.
In these cases there could be contributions to the final quantity used for the trigger decision from more than one \pv, thus multiple recorded collisions within the \rbc{} are biased by that decision.
The ability to identify and remove the triggering collision(s) within a bunch crossing is necessary to produce a trigger-unbiased dataset, and this cannot be done for such triggers based on vertex-independent quantities, thus they are not considered further.
It is easiest to identify the triggering collision(s) when using signatures involving charged particles, which can be matched to the vertex they originate from.
The trigger selections with the largest acceptance rate during Run~2 were single-electron and single-muon signatures.
These triggers amount to between one quarter to one third of the total recorded dataset, and additionally provide a straightforward intepretation of the triggering vertex.
For that reason, only \bcs{} recorded by one of these selections are considered.

In order to identify the triggered process in each bunch crossing, reconstructed electrons and muons are matched to their trigger-reconstructed counterparts. The set of trigger electrons and trigger muons, which are associated with having satisfied the single-electron or single-muon triggers, are collected. These provide a handle to understand which physics object(s) in the \rbc{} was or were responsible for satisfying the trigger requirement for the \rbc{}. For a given electron or muon, the closest reconstructed particle of the same type within this set of trigger electrons or trigger muons is selected, and is considered matched if they are at a distance of $\Delta R \leq 0.1$. Once an electron or muon is matched to the corresponding triggering object, the collision it originates from is determined by associating the electron or muon's primary inner detector track to a \pv{} in the event. In the case of electrons (muons), this is done by requiring the significance of the transverse impact parameter of its track be $|d_0/\sigma_{d_0}| < 3$ (5), and that the longitudinal distance to the \pv{} be $|\Delta z_0\sin\theta| < 0.5\text{ mm}$. The \pv{} thus associated to this trigger-matched electron or muon is then removed from the \rbc{}, and the remaining \ppvs{} are used to construct the pile-up dataset. In \rbcs{} with more than one triggering lepton, all \pvs{} that are similarly trigger-matched to one or more triggering electrons or muons are removed.

\begin{figure}[t]
\centering
\includegraphics[width=0.48\textwidth]{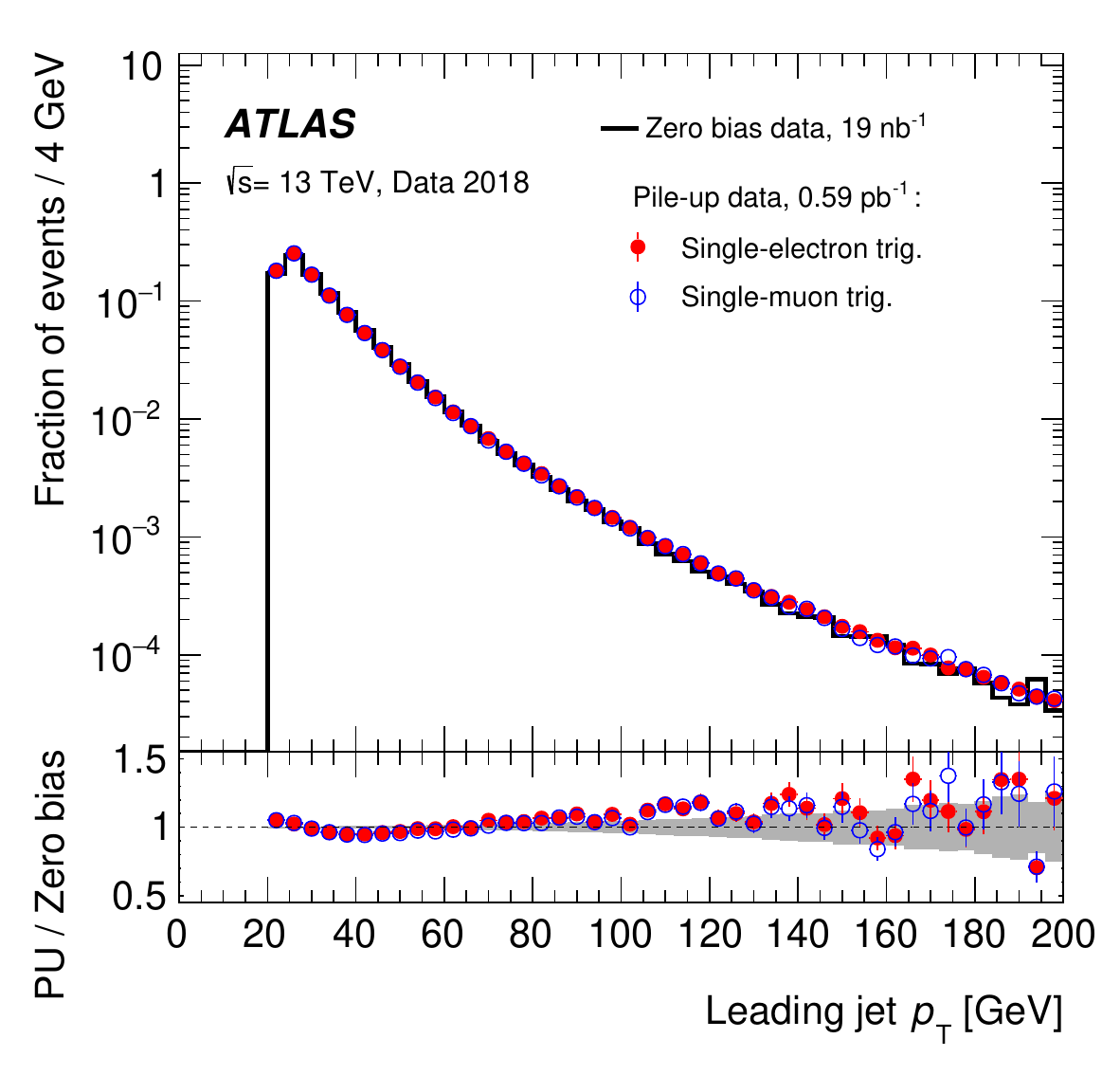}
\caption{Comparison of the leading jet \pT{} between the pile-up reconstructed dataset and the reference zero bias dataset as collected in 2018.
The pile-up dataset is split between the \rbcs{} collected with a single-electron trigger (filled circles) from those recorded with a single-muon trigger (open circles). The selection discussed in \Section{sec:PV-selections} is applied.
}
\label{fig:jetPtTrigger}
\end{figure}

As a way of validating the correct removal of the \tpv{} from the \rbc{}, the pile-up dataset is compared to the zero bias dataset described in \Section{sec:datasamples}. The same jet reconstruction and jet-vertex association approach discussed in \Section{sec:reco} is applied to both pile-up and zero bias datasets. Figure~\ref{fig:jetPtTrigger} shows the distribution of the leading jet \pT{} for pile-up, separated by whether the \rbcs{} were selected with the single-electron or single-muon trigger requirement, and for zero bias data.
The pile-up and zero bias results all make use of the dijet selection discussed in \Section{sec:PV-selections} in order to filter the list of vertices under study.
The pile-up and zero bias data agree within their corresponding statistical uncertainties for most of the spectrum. There are some residual differences between the pile-up and zero bias datasets for jet \pT{} above roughly $100\,\GeV{}$, but this is beyond the primary regime of interest for the pile-up dataset; even in this regime, there is great consistency between the single-electron and single-muon triggered \rbcs.

\subsection{Primary vertex selections}
\label{sec:PV-selections}

When constructing the pile-up dataset, the contribution from inter-dependent physics events within the same \rbc{} should be reduced as much as possible. With this aim, a set of quality criteria are designed and applied sequentially, first to all jets in the \rbc{} and then to the \ppvs{}.

After defining the set of jets to use from a given \pv{}, the \ppvs{} are selected according to the objects associated to them. Firstly, the \tpvs{}, together with any other \ppvs{} within $|\Delta z(\text{TPV, PPV})| < 2$~mm of a \tpv{}, are vetoed, leaving the remaining \pvs{} in the \rbc{} unbiased by the trigger.
The 2~mm requirement is used to avoid the \tpv{} contributing to \ppv{} jet reconstruction: charged PFOs are matched to the closest \pv{} within 1~mm, and thus a 2~mm window removes the region of possible confusion.
If at least one \tpv{} could not be found for a given \rbc{}, due to inefficiencies in the lepton reconstruction, trigger-matching procedure or lepton-vertex association, then the whole \rbc{} is vetoed.
The main physics process populating this dataset is dijet production, and therefore the remaining \ppvs{} are only considered if there are at least two jets matched to the \ppv{}. In order to reduce the contribution from overlapping high-momentum signals in the detector, the two leading jets from the \pv{} are required to have $\fCPV=1$. Additionally, out-of-time pile-up signals can lead to jets with different energy if they overlap with one of the in-time signals in the calorimeter. Therefore, a jet timing\footnote{The jet timing is calculated as the energy-squared-weighted sum of the timing of the constituent clusters, which in turn is the energy-squared-weighted sum of the timing of the cells within each cluster.} requirement of 10~ns is applied on the two leading jets in the \pv{} to remove overlaps with signals from neighbouring \bcs~\cite{PERF-2016-06}.

\begin{figure}[t]
\centering
\includegraphics[width=0.48\textwidth]{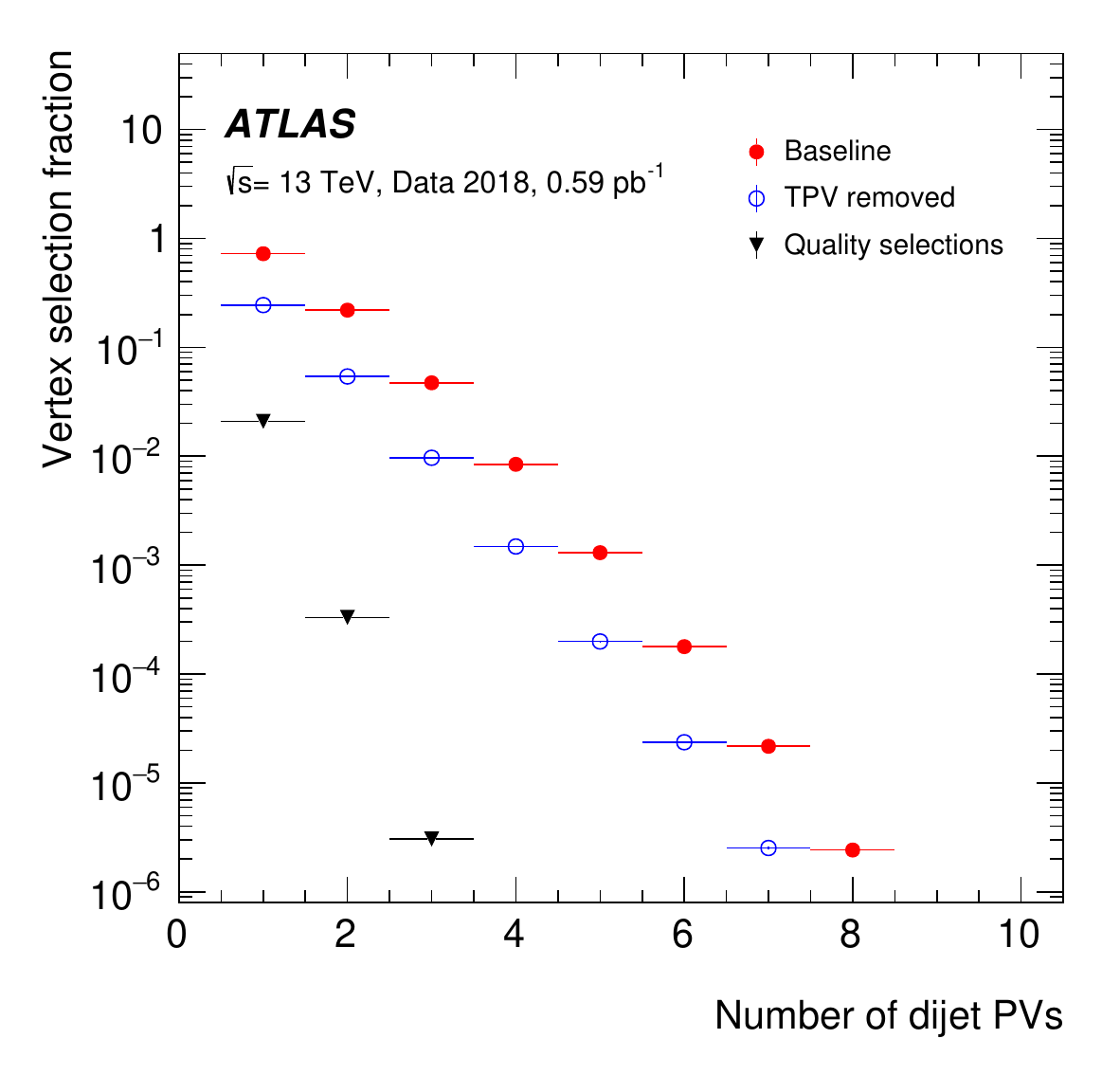}
\caption{Efficiency of the number of reconstructed \ppvs{} for different selections. ``Baseline'' (filled circles) indicates that the \rbc{} must have satisfied either the single-electron or single-muon trigger, and a loose selection of at least two jets with $\pT{}>15\,\GeV{}$ and satisfying JVT is applied; ``\tpv{} removed'' (open circles) stands for the ``baseline'' selection plus the additional requirement of a \tpv{} being found and removed, and ``Quality selections'' (filled triangles) stands for the set of selections introduced in~\ref{sec:PV-selections} such as only keeping vertices with two associated jets. Statistical uncertainties are included for all selections, and are negligible.
}
\label{fig:NPVDistributionEfficiency}
\end{figure}

The distribution of the number of \pvs{} that survive all these selections within each \bc{}, and which thus are used in subsequent studies, is shown in \Figure{fig:NPVDistributionEfficiency}.
The efficiency of correctly finding the \tpv{} and removing it is roughly $1/3$, which is determined by inefficiencies in the online-offline lepton matching algorithm together with inefficiencies in the lepton-vertex association strategy used.
This efficiency does not depend on the other activity in the \rbc{}, and therefore the distribution of the number of \pvs{} is not significantly sculpted after the \tpv{} removal.
The efficiency of the quality selections that are applied for the final pile-up dataset is about $1/10$ for \rbcs{} with a single dijet \ppv{} with respect to the \tpv{} removal step.
These more stringent selections are now applied to each \ppv{}, resulting in the distribution of number of \pvs{} falling much faster, where the proportion of \rbcs{} with four good dijet \ppvs{} is found to be one in a billion, and no \rbcs{} were found in the full 2018 dataset with five good dijet \ppvs{}.
The total efficiency of having at least one \ppv{} satisfying all selections in a given \bc{} is $(2.1220\pm0.0006)\%$, considering only the statistical uncertainty, over the full 2018 pile-up dataset.

\subsection{Comparing the behaviour of jets from different vertices}
\label{sec:other-BC-activity}

At this stage, the following effects have been mitigated: hard-scatter jets overlapping with underlying event (\Section{sec:reco:calib}), hard-scatter jets overlapping with other hard-scatter jets from a separate collision (\Section{sec:jet-jet-overlaps}), removal of triggering collisions (\Section{sec:trigger-bias}), and the selection of high-quality \pvs{} (\Section{sec:PV-selections}).
The requirement of $\fCPV{}=1$ is further tightened: any \pv{} with a jet failing this requirement is vetoed, thus only \pvs{} where there is no double-use of calorimeter energy deposits are retained.
At this point the properties of jets reconstructed with respect to different \pvs{} may be compared, with the expectation that the properties of a jet should not depend on which \pv{} the jet comes from.
A useful quantity for this purpose is the jet charged fraction, $\sum \pT^\mathrm{tracks} / \pT^\mathrm{jet}$, where the sum runs over all tracks consistent with the \pv{} of interest that are ghost-associated with the jet.
This quantity mixes both tracking and calorimeter information, with the numerator robust against other activity in the same \bc{}, and the denominator dependent upon a proper handling of calorimeter contributions.
It is also an important variable for other reasons: the jet charged fraction is one of the variables used in the jet calibration procedure~\cite{JETM-2018-05}.

\Figure{fig:jetChargedFractionPVs} shows the charged fraction of pile-up jets from the leading and sub-leading \ppvs{} in the \rbc{}, in \rbcs{} where there are at least two \ppvs{}, after a selection of $50\,\GeV{}^2 < \pTvtx{} < 150\,\GeV{}^2$ on each of the \ppvs{}. This is required to mitigate the implicit vertex ordering bias, as introduced in \Section{sec:reco:calib}.
Excellent agreement is observed, demonstrating that the properties of the reconstructed jets are indeed independent of the \ppp{} collision from which they originate.
The small residual difference goes in the expected direction, and could be further reduced by tightening the selection on \pTvtx{}.
The potential dependence of jets originating from \ppvs{} on the presence of other much higher energy collisions in the same \rbc{} is not typically considered in a \pv{}0-based analysis strategy, but is important to consider when using the pile-up dataset.
\Figure{fig:jetChargedFractionTPVBins} studies such a possibility through comparisons of the charge fraction of pile-up jets for three different bins of \tpv{} \pTvtx{}, where the most energetic bin considered requires at least twenty times more \pTvtx{} for the \tpv{} as opposed to the \ppv{}; the charged fraction is robust against such additional energy in the detector, showing that the overlap removal and calibration procedures are working well.

\begin{figure}[t]
\centering
\subfigure[]{\includegraphics[width=0.48\textwidth]{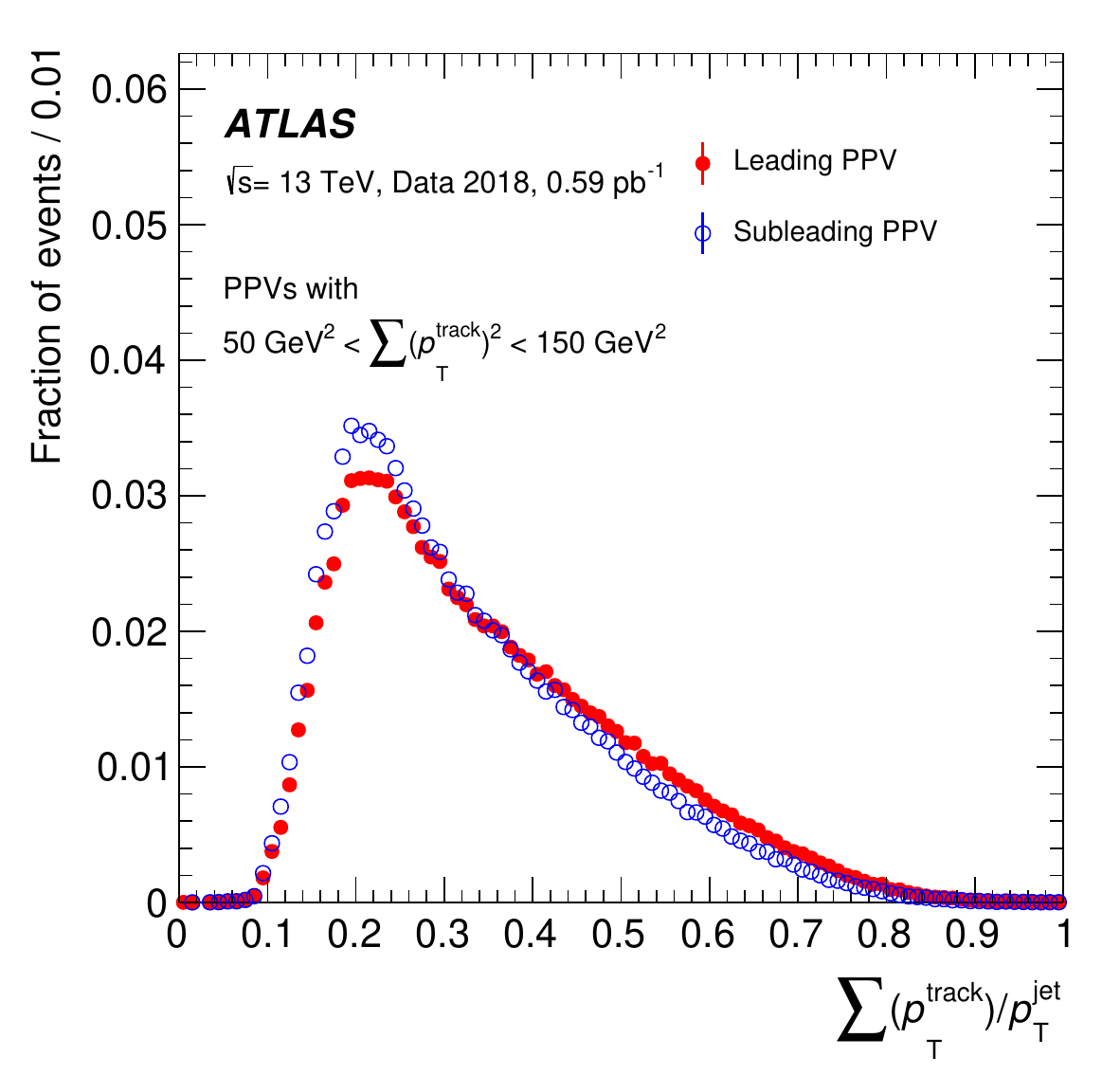}\label{fig:jetChargedFractionPVs}}
\subfigure[]{\includegraphics[width=0.48\textwidth]{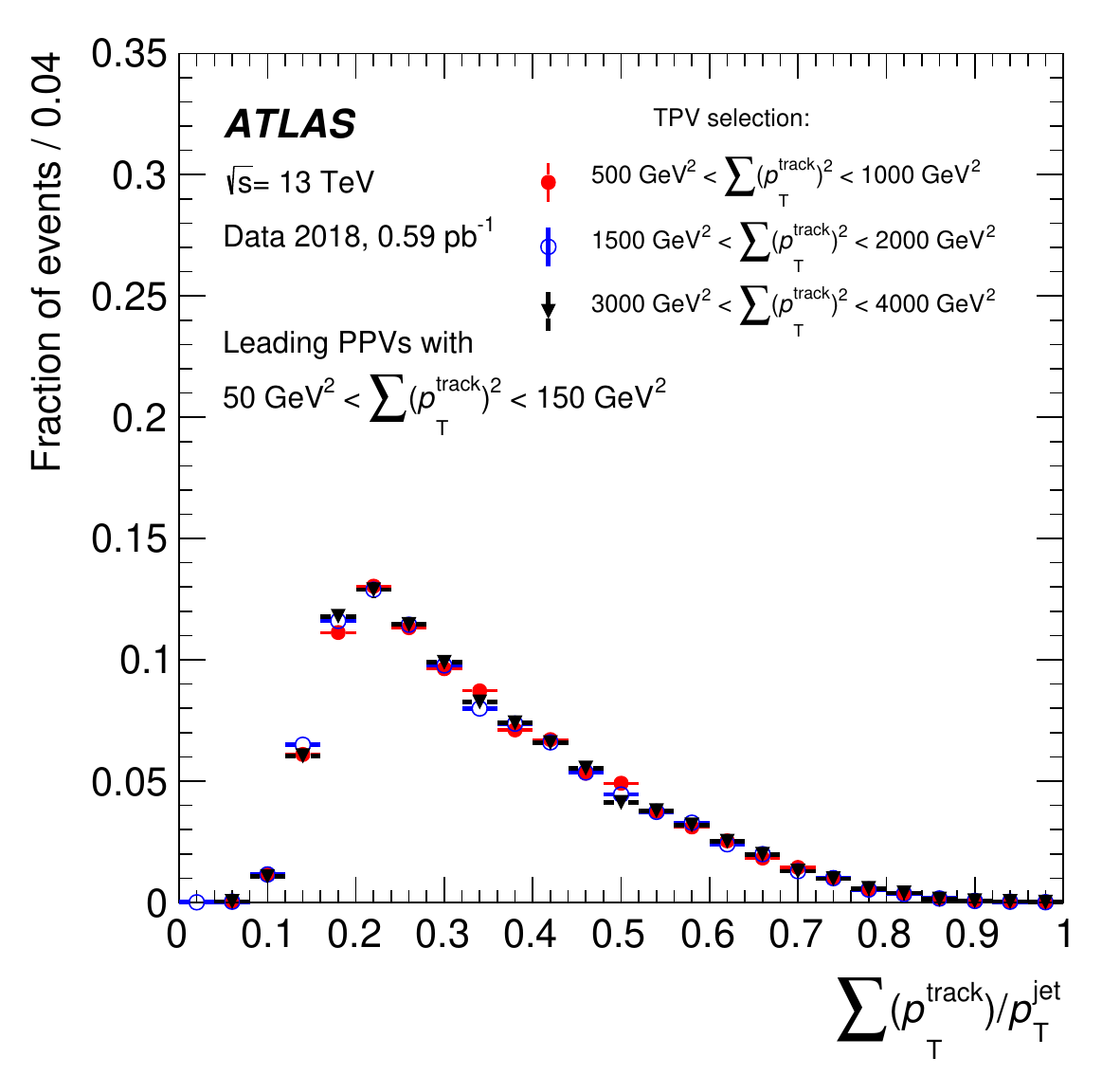}\label{fig:jetChargedFractionTPVBins}}
\caption{Jet charged fraction for (a) jets reconstructed with respect to the leading and sub-leading \ppv{}, and (b) the leading \ppv{} from \rbcs{} with \tpvs{} at different energy scales, for the full 2018 pile-up dataset. A minimal selection of at least two jets with $\pT>20$\,\GeV{} and satisfying JVT is applied. Statistical uncertainties are included for all selections, and are negligible.
}
\end{figure}

\subsection{Visualising the pile-up dataset}
\label{sec:reco:BC}

Following the reconstruction of the individual objects within a given \rbc{}, and the classification and treatment of the collisions within the \rbc{} (such as \tpvs{} vs \ppvs{}), it is beneficial to also consider the \rbc{} as a whole.
In this way, it is possible to cross-check that the reconstruction is working as intended, and that the same detector signal is not being used for multiple different physics objects.

In order to support such studies, it is convenient to visualise the entire \rbc{}, such as through the use of a 2D ($y,\phi$) projection, where $y$ is the rapidity.
\Figure{fig:ED_2D} shows a pair of such visualisations: one is an example of a particularly useful \rbc{}, with a di-muon \tpv{} and two good dijet \ppvs{}, and the other is of a problematic \rbc{}, where jets from different \ppvs{} overlap and are therefore both rejected by the $\fCPV{}=1$ requirement.
A full visualisation of the same useful \rbc{}, at a different angle and with the detector geometry superimposed, is provided in \Figure{fig:ED_3D}.

\begin{figure}
\centering
\subfigure[]{\includegraphics[width=0.75\textwidth]{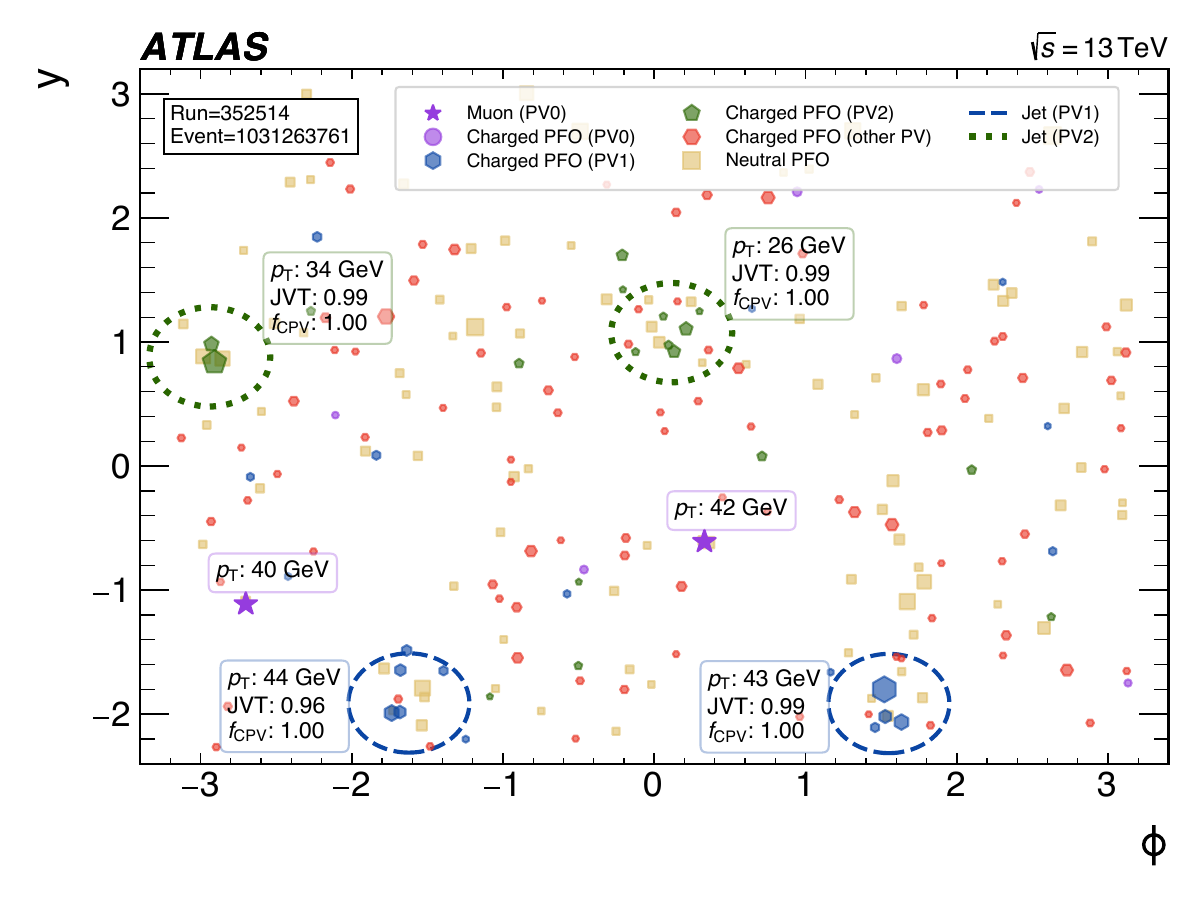}\label{fig:ED_2D_good}}\\
\subfigure[]{\includegraphics[width=0.75\textwidth]{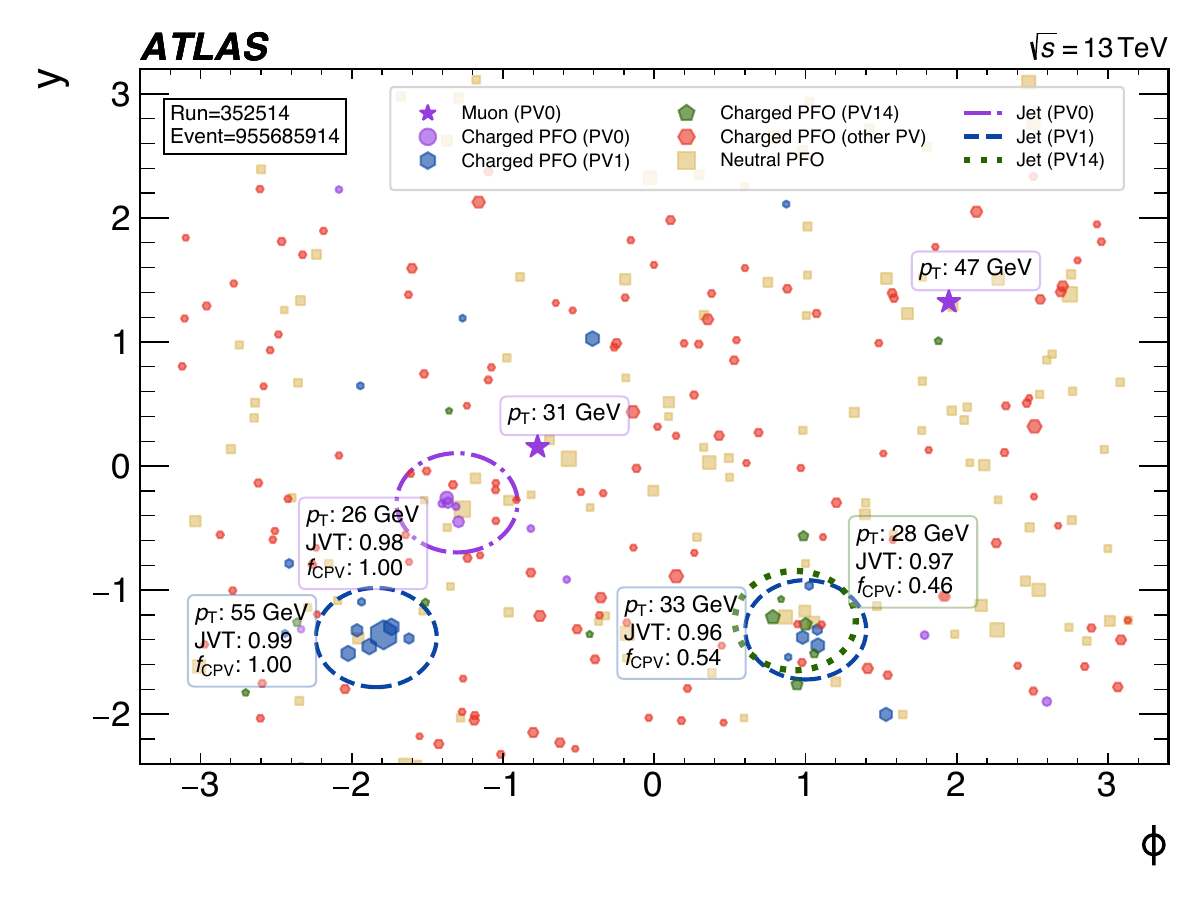}\label{fig:ED_2D_bad}}
\caption{
A ($y,\phi)$ visualisation of the entire recorded bunch crossing for (a) a muon-triggered \rbc{} with a di-muon \tpv{} and two good dijet \ppvs{}, and (b) a muon-triggered \rbc{} with a pair of \ppvs{} with overlapping jets.
In addition to listing the \pT{} of each physics object, jets also provide information on variables used to quantify the degree of consistency of the jet originating from the current \pv{} interpretation (JVT), and to quantify the degree to which the jet in question is isolated from other jets ($f_\mathrm{CPV}$).
The invariant mass of the di-muon system in (a) is 84\,\GeV{}, and the \pT{} is 5\,\GeV{}.
The Z boson candidate in (b) is more energetic, with an invariant mass of 88\,\GeV{} and a \pT{} of 23\,\GeV{}; in this case, the Z boson candidate is also balanced by a jet.
}
\label{fig:ED_2D}
\end{figure}

\begin{figure}
\centering
\includegraphics[width=\textwidth]{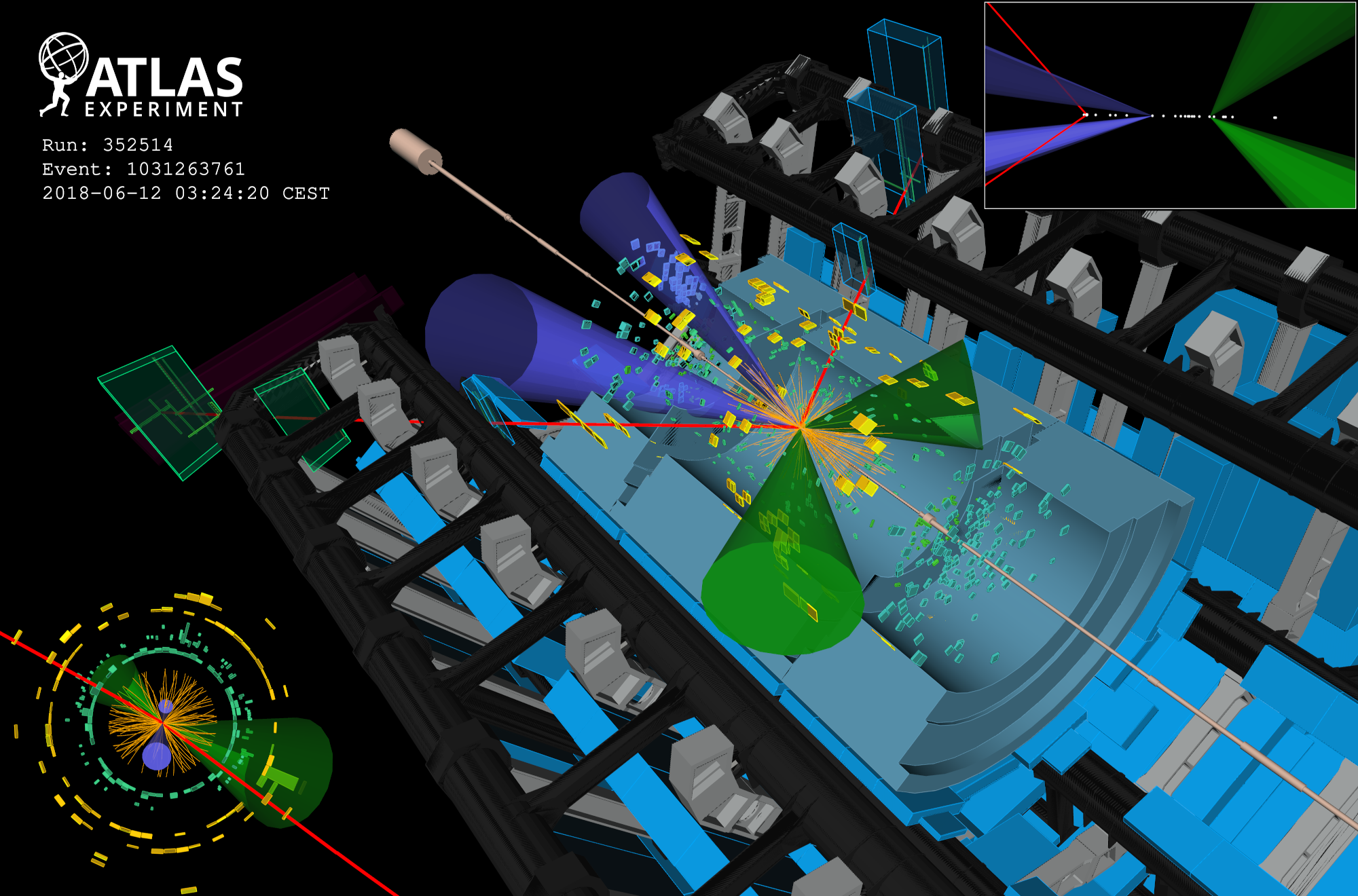}
\caption{
An event display of a muon-triggered recorded bunch crossing from data collected on the 12th of June 2018 (Run 352514, Event 1031263761), shown together with the ATLAS detector geometry and highlighting the reconstruction of pile-up collisions. The bunch crossing contains a di-muon event, consistent with a Z boson, as the triggering primary vertex and two other good dijet pile-up primary vertices.  The two muons are shown as red lines.  The jets are colour-coded to show which vertex they come from: the two blue jets come from a second vertex, and the two green jets come from a third vertex. This is highlighted in the close-up to the beamspot shown in the top right corner, where the pair of muons and the two pairs of jets all come from clearly separated primary vertices. This is compatible with them being produced in independent \ppp{} collisions within the same \bc{}. Tracks are shown in orange, the muon chambers associated to the muon tracks are shown as blue and green boxes, and the green and yellow blocks correspond to energy depositions in the electromagnetic and hadronic calorimeters, respectively. The view in the bottom left shows the plane transverse to the beam direction.
}
\label{fig:ED_3D}
\end{figure}


%
\section{Extraction of the jet energy resolution using the pile-up dataset}
\label{sec:JER}

%

The jet energy resolution (JER) quantifies the extent to which the detector response to a given jet can be known, and thus is a measure of the ability to differentiate between distinct calibrated signals.
The JER is an important experimental quantity for many measurements of SM processes and searches for new phenomena.
The JER is typically parameterised using a functional form inspired by the calorimeter resolution, containing terms accounting for noise ($N$), stochastic ($S$), and constant ($C$) contributions~\cite{Wigmans:2017vgs}:
\begin{equation}
\frac{\sigma(\pT{})}{\pT{}} = \frac{N}{\pT{}} \oplus \frac{S}{\sqrt{\pT{}}} \oplus C .
\label{eqn:JER}
\end{equation}

One particularly powerful means of measuring the JER is to exploit the abundant dijet production cross-section, together with the conservation of transverse momentum, in \ppp{} collisions at the LHC\@.
A true dijet event should consist of a pair of perfectly balanced jets, and thus any difference in the \pT{} of the two jets would be attributed to the detector resolution.
In practice, perfect dijet events cannot be guaranteed, thus imbalances in the \pT{} of the pair of jets in a dijet system may also be related to the presence of additional radiation in the event.
Analysis selections can be used to suppress, or quantify the impact of, such additional radiation in the event; however, these selections also can cause biases, and have their own uncertainties.
Despite these caveats, measurements of imbalances in dijet events are still one of the best ways to extract the JER at the LHC.

\subsection{Motivation for using the pile-up dataset}
\label{sec:JER_motivation}

The ATLAS Run~2 measurement of the JER relied heavily upon such dijet balance techniques to determine the values of the stochastic and constant terms, while a separate method, referred to as the random cone technique, was used to constrain the noise term~\cite{JETM-2018-05}.
While the dijet balance approach is sensitive to the noise term through a combined fit to all three JER parameters, the random cone technique measures the noise term directly.
The random cone procedure has worked well so far, but it relies on the detector responding to stochastic energy deposits irrespective of the presence or absence of other activity: the response should not depend on whether a jet overlaps with the stochastic energy deposit, or not.
This has not yet been observed to be an issue, but as pile-up increases the noise thresholds will also increase, and thus stochastic energy will increasingly only survive if it overlaps with a jet.
More advanced jet definitions, including constituent-level pile-up mitigation methods such as constituent subtraction~\cite{Berta:2014eza,Berta:2019hnj} and SoftKiller~\cite{Cacciari:2014gra}, also change the behaviour of stochastic contributions inside and outside of jets; such methods can therefore impact the applicability of the random cones procedure.
It is therefore useful to be able to extend the dijet method to have increased sensitivity to the noise term, thereby providing a cross-check of the random cones method.

The dijet balance approach as presented in Ref.~\cite{JETM-2018-05} has limited sensitivity to the noise term due to data-related statistical uncertainties growing quickly at low \pT{}, and simulation-related systematic uncertainties reducing the weight of the dijet balance in the combined JER measurement.
The knowledge of the relevant systematic uncertainties has improved since that result~\cite{JETM-2022-01,ATL-PHYS-PUB-2022-021}, but the dataset is the same, thus the statistical uncertainty at low \pT{} is unchanged.
This is where the pile-up dataset can play a role: the pile-up dataset can provide the abundant source of low-\pT{} dijet events required to extend the sensitivity to the noise term.
Such a measurement is also an excellent means of demonstrating the validity of the pile-up dataset, as any mistakes in reconstructing or selecting pile-up collisions will spoil the expected \pT{} balance of dijet events, thus showing up as a degraded JER.

A comparison of the statistical power of the single-jet-triggered dataset, as used in previous publications such as Ref.~\cite{JETM-2018-05}, and the pile-up dataset is provided in \Figure{fig:JER_numEvents}.
This plot shows the number of selected events in the pile-up-based and trigger-based datasets as a function of $\pTavg{} = \frac{1}{2}(\pT^{j_1} + \pT^{j_2})$, and for bins defined based on the range of usage of each single-jet trigger.
This binning was chosen to support a direct statistical comparison of the pile-up-based and single-jet-trigger-based approaches; it is not optimal for extracting the JER, and thus it does not correspond to the binning used in the rest of this Section.
The pile-up dataset falls smoothly, as it is a single inclusive dataset, while the single-jet triggered dataset varies with the rate of the prescaled triggers used.
The single-jet triggered dataset is further suppressed at low \pTavg{} by the selection requirements: the selection only considers \pv{}0, and at low \pTavg{} it is rare for the leading vertex to produce jets at such a low energy scale.
The results show that in 2018 data, the pile-up dataset improves the statistical precision of the JER extraction for $\pTavg{} < 65\,\GeV{}$, with 50 times more good dijet events for $\pTavg{}\in[35,45]\,\GeV{}$ and 10 times more good dijet events for $\pTavg{}\in[45,65]\,\GeV{}$.
Each of these single-jet triggers had a typical rate slightly larger than 4\,Hz in 2018 data taking. Combining the first two bins in \Figure{fig:JER_numEvents}, the pile-up dataset provides the statistical equivalent of more than 240\,Hz inclusive single-jet trigger after applying the dijet JER selection criteria.
Adding such a 240\,Hz single-jet trigger to the ATLAS trigger strategy is not possible within the constraints and overall physics goals of the experiment: such a trigger would have had a higher rate than either the single electron or single muon triggers, and would have represented roughly 1/5 of the full data-taking rate in 2018.
The ability to extract such a statistically abundant dataset, without any impact on the trigger, data acquisition, or storage systems, is thus a major success of the pile-up-based data analysis methodology.

\begin{figure}
\centering
\includegraphics[width=0.48\textwidth]{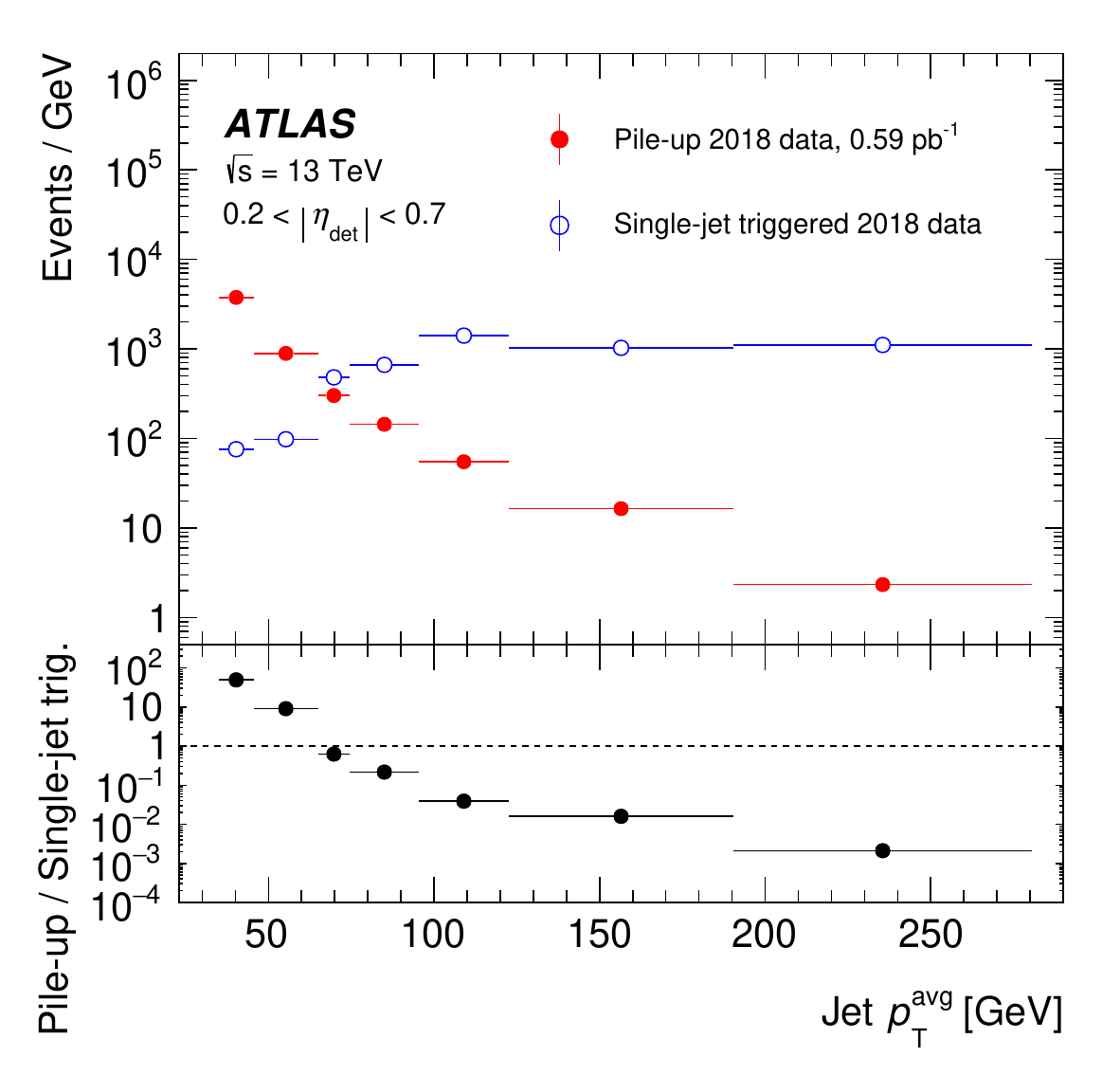}
\caption{
The number of events satisfying the selections used for the dijet JER measurement within the reference region $(0.2 < |\eta_\mathrm{det}| < 0.7)$, for the pile-up dataset (filled markers) and the single-jet triggered dataset (open markers).
The bin ranges are chosen to match the range over which each single-jet trigger is used: the lower edge is the point at which the trigger becomes fully efficient, and the upper edge is the point at which the next trigger becomes fully efficient and thus starts to be used.
Statistical uncertainties are included, and are negligible.
The ratio panel facilitates an easy determination of the range over which a given dataset is statistically favoured.
The integrated luminosity for each single-jet trigger shown here can be found in \Figure{fig:lumi_pile-upVsTrig_2018}.
}
\label{fig:JER_numEvents}
\end{figure}

\subsection{Dijet asymmetry in pile-up collisions}
\label{sec:JER_details}

The calculation of the JER in dijet events follows the method used for the ATLAS Run~2 nominal JER results~\cite{JETM-2018-05}.
The dijet asymmetry is defined as:
\begin{equation}
\mathcal{A} \equiv \frac{\displaystyle \pT^\mathrm{probe} - \pT^\mathrm{ref}}{\pTavg{}},
\end{equation}
where $\pT^\mathrm{ref}$ is a jet in the reference region of the detector, defined as $0.2<|\eta_\mathrm{det}|<0.7$; this is the region where the calorimeter is the most uniform, thus providing a clean measurement of the JER\@.
The probe jet can then be used to study the JER in either the same or other $\eta$ regions, but this demonstrative result only considers jets where the probe is also within the reference region.
In this simplified case, the JER can be extracted from the standard deviation of the asymmetry distribution, $\sigma_\mathcal{A}$, as:
\begin{equation}
\left\langle\frac{\sigma_{\pT{}}}{\pT{}}\right\rangle = \frac{\sigma_\mathcal{A}}{\sqrt{2}}.
\end{equation}

This provides the complete resolution, but the quantity of interest is rather the detector resolution, which requires the removal of particle-level effects.
The same procedure as Ref.~\cite{JETM-2018-05} is used to correct for particle-level effects, whereby the asymmetry of the distribution is fit using a convolution of a particle-level-derived function and a Gaussian function that represents the detector resolution.
The \PYTHIA8 samples described in \Section{sec:datasamples} are used to fit the particle-level asymmetry.
The procedure does not need to be adapted for the use of the pile-up dataset, as the particle-level representation of a jet is independent of which \pv{} it happens to originate from; the act of identifying primary vertices is a reconstruction-level procedure.

The analysis selections for the extraction of the JER are also similar to Ref.~\cite{JETM-2018-05}, including a requirement of $|\Delta\phi(j_1,j_2)|>2.7\,\mathrm{rad}$ in order to enforce a back-to-back dijet balance topology, with an additional veto on third jet radiation of $\pT^{j_3} < \max\left(25\,\GeV{},0.25\cdot\pTavg{}\right)$.
However, the third-jet radiation veto was modified in two ways.
First, the absolute \pT{} threshold used for the third-jet radiation veto was lowered from 25\,\GeV{} to 15\,\GeV{} within the central region ($|\eta_\mathrm{det}| < 2.4$), leaving the relative veto threshold unchanged at $0.25\cdot\pTavg{}$.
The objective of this effort is to study lower \pTavg{} regions, and thus the third-jet veto threshold must also be lowered to retain a similar level of certainty that the asymmetry of the dijet system is not due to the presence of additional radiation; this was therefore applied to both the pile-up dataset and the single-jet triggered dataset.
This change was only applied to the central region because such low \pT{} jets suffer from significant fake rates in the forward region due to the lack of track information, whereas in the central region the use of particle flow and JVT is generally sufficient to ensure the jet is from the collision under study.
Second, instead of removing only a single collision, the pile-up dataset removes the entire bunch crossing if there is a forward jet ($|\eta_\mathrm{det}| \geq 2.4$) satisfying the third-jet \pT{} threshold of 25\,\GeV{}.
This is important because the forward region has no tracking information, and thus it is not possible to achieve sufficient certainty of which vertex the forward jet originates from; it is therefore safer to remove the entire bunch crossing, rather than risk biasing the asymmetry measurement.
As a summary, the additional radiation suppression selections for the pile-up dataset are:
\begin{itemize}
\item Require individual \ppvs{} to satisfy $|\Delta\phi(j_1,j_2)|>2.7\,\mathrm{rad}$;
\item Require individual \ppvs{} to satisfy $\pT^{j_3} < \max\left(15\,\GeV{},0.25\cdot\pTavg{}\right)$ for jets with $|\eta_\mathrm{det}|<2.4$;
\item Require entire \rbcs{} to satisfy $\pT^\mathrm{jet} < \max\left(25\,\GeV{},0.25\cdot\pTavg{}\right)$ for jets with $|\eta_\mathrm{det}|\geq2.4$.
\end{itemize}

In addition to these updates to the analysis-level dijet JER selections, the pile-up dataset requires a series of selections to ensure that only good \ppp{} vertices are used, as discussed in \Section{sec:PV-selections}; these selections are not required for the single-jet triggered approach, where only \pv{}0 is used.
Even though the JER applies selections based on the first three jets, the requirement of $\fCPV{}=1.0$ for only the two leading jets is conservative: overlapping jets will have a \pT{} that is inflated with respect to the reconstruction of only a single vertex, and thus will trigger the third-jet veto at a point where the effective \pT{} imbalance is smaller than for a non-overlapping jet.
The application of these vertex selection criteria are shown in \Figure{fig:JER_asym_byStep}, where the act of removing the \tpv{} is shown to have the largest impact on the asymmetry.
This is expected, given that the \tpv{} removal is necessary to make the dataset unbiased with respect to the triggering process, but the asymmetry plot is a useful confirmation.
The asymmetry distribution after the full set of selections is then compared for the pile-up and single-jet triggered datasets in \Figure{fig:JER_asym_overall}, showing good agreement, and demonstrating the dramatically increased statistical precision of the pile-up dataset at low \pTavg{}.

\begin{figure}
\centering
\subfigure[]{\includegraphics[width=0.48\textwidth]{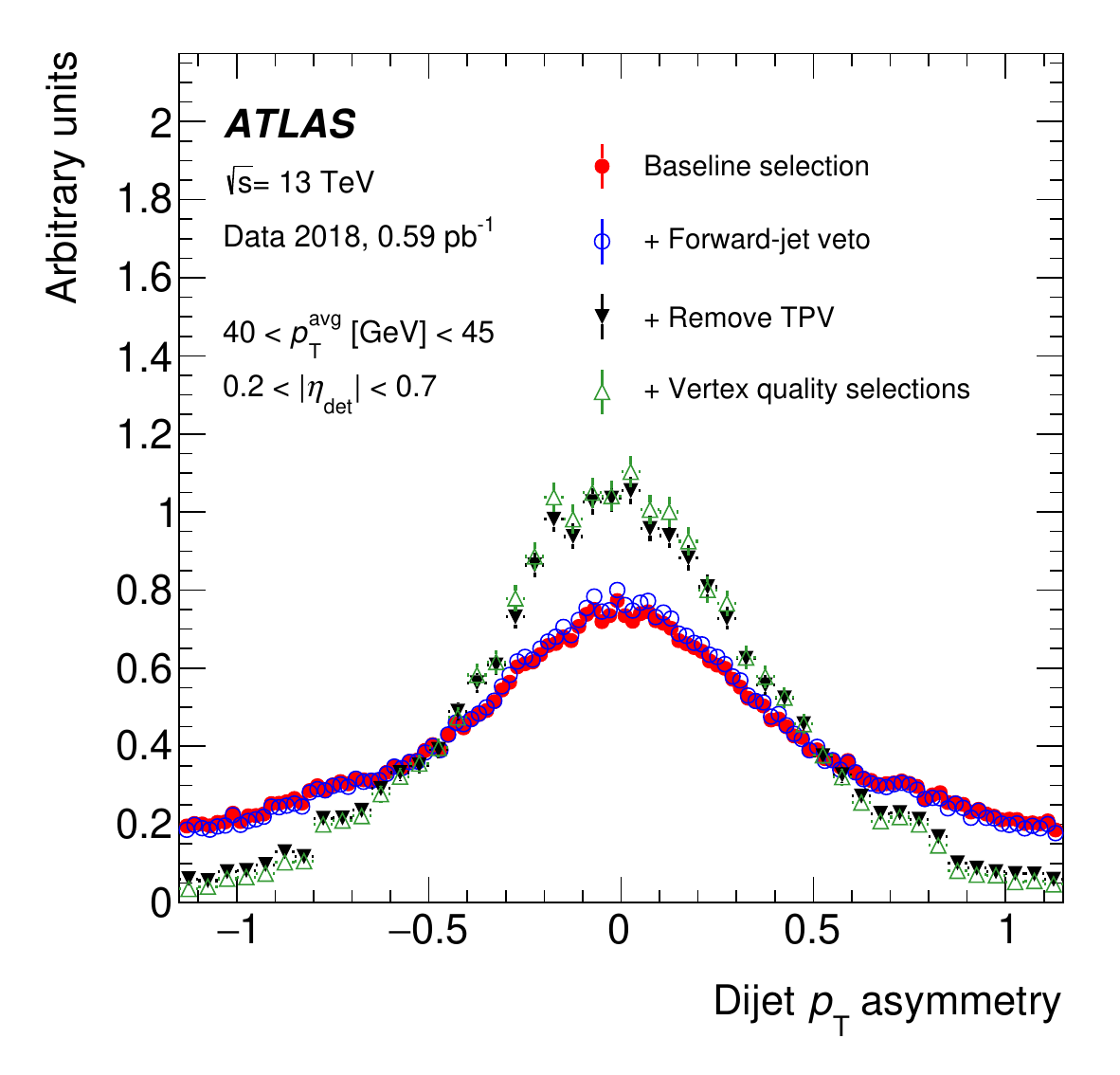}\label{fig:JER_asym_byStep}}
\subfigure[]{\includegraphics[width=0.48\textwidth]{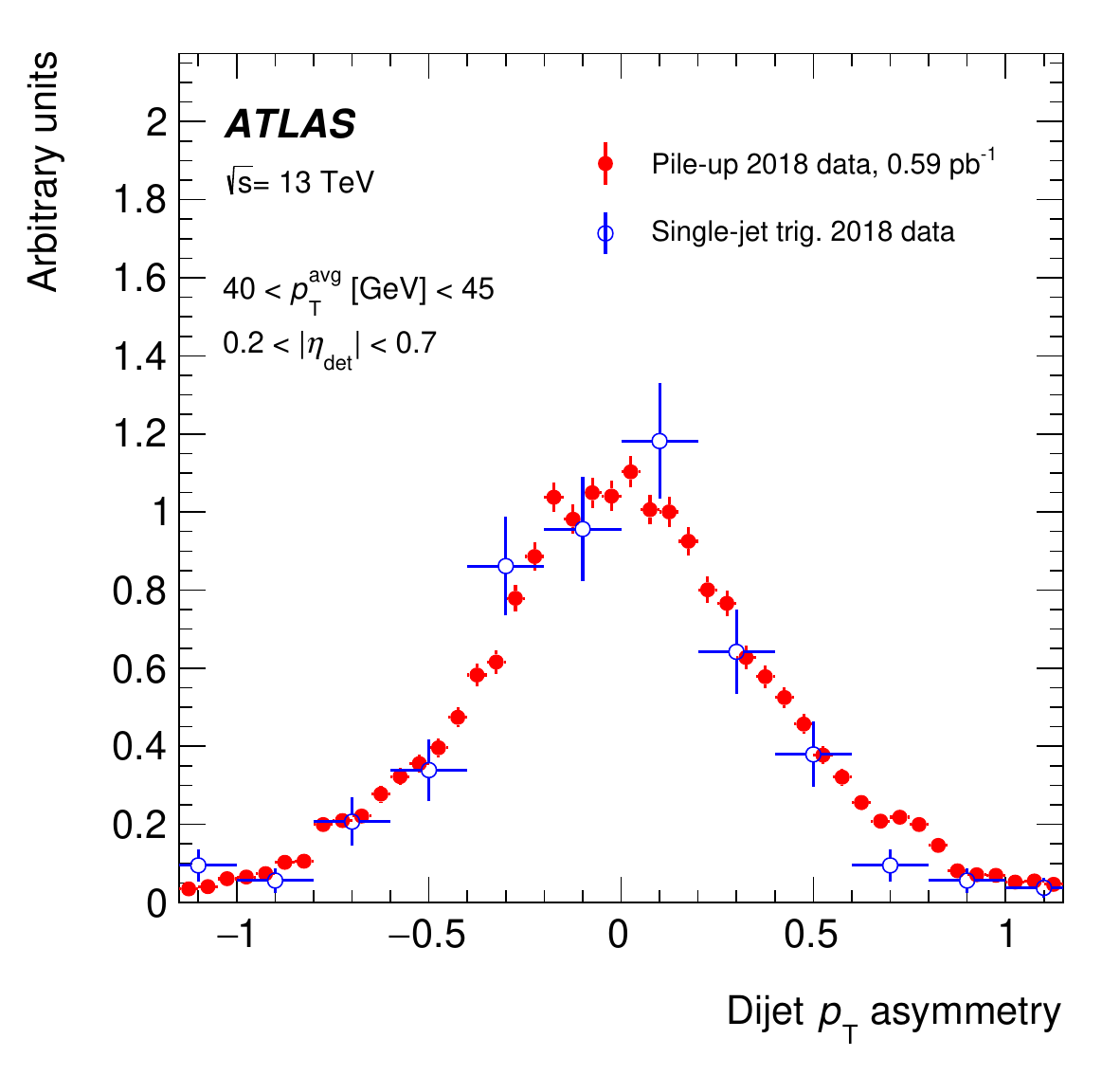}\label{fig:JER_asym_overall}}
\caption{
A comparison of the dijet asymmetry distribution for (a) the pile-up dataset, following a sequential set of vertex selection criteria, and (b) the pile-up dataset in filled points, as compared to the single-jet triggered dataset in open points.
Both of the plots are shown using the corresponding full 2018 dataset.
``Vertex quality selections'' stands for the set of selections introduced in \Section{sec:PV-selections} such as only keeping vertices with two associated jets.
The asymmetry fit range used to extract the JER for this \pTavg{} bin is from -0.5 to 0.5.
}
\label{fig:JER_asym}
\end{figure}

The lowest bin, $\pTavg{}\in[35,40]\,\GeV{}$, was chosen such that the subleading jet \pT{} remains above 20\,\GeV{} across the full asymmetry fit range.
Given that the pile-up dataset increases in statistical power as the \pTavg{} value decreases, it would be interesting to study the use of this approach for even lower \pTavg{} values, but that would require corresponding improvements in the jet energy scale calibration in order to not be overwhelmed by systematic uncertainties in this regime.

Asymmetry measurements were performed separately for the 2017 and 2018 datasets, for all of the \pTavg{} bins considered, in order to cross-check whether there are any unexpected time-dependent effects that would have been missed when only looking at 2018 data.
Good agreement was observed for all \pTavg{} bins, thus the two years were combined taking the integrated-luminosity-weighted mean of the resulting JER values.

\subsection{Results}
\label{sec:JER_results}

In order to use dijet events to extract the JER, using the NSC function as defined in Eq.~\eqref{eqn:JER}, it is necessary to measure the JER via the dijet asymmetry over a wide range of \pTavg{}.
The single-jet-triggered dataset was used in Ref.~\cite{JETM-2018-05} to do this, the results of which are shown together with the JER as measured using the pile-up-based approach in \Figure{fig:JER_dijet}; both results use only the data recorded in 2017 for this study.
The different results agree well over the full range considered, with the exception of one bin at $\pTavg\approx65\,\GeV{}$, which shows some tension.
This was already the most discrepant point in Ref.~\cite{JETM-2018-05}, but the tension is further visually enhanced due to the single-jet-triggered result appearing to be an under-fluctuation, while the pile-up result appears to be an over-fluctuation.
The bin in question is challenging for the single-jet-triggered approach due to the presence of a threshold effect: the JVT requirement in the single-jet-triggered analysis is only used for $\pT^\mathrm{jet}<60\,\GeV{}$, and jets in this bin can be either below or above this threshold.
The fluctuation in the pile-up-based result appears to be statistical in nature: the use of 2018 data leads to a lower JER for this specific bin, where the value observed in 2018 pile-up data is roughly halfway between the 2017 single-jet-triggered and 2017 pile-up data results.

A comparison of the overall trends seen for the two approaches in \Figure{fig:JER_dijet} shows the trade-off: the single-jet triggered dataset is relatively insensitive at low \pT{}, and thus has limited sensitivity to the noise term, while the pile-up dataset quickly runs out of statistics at high \pT{}, and thus is not useful for measuring the constant term.
It is therefore a natural next step to combine these two datasets to obtain an improved sensitivity across the largest possible \pTavg{} range, thereby maximising the ability to extract the noise, stochastic, and constant terms.
A simplified statistics-only combination of the pile-up-based and single-jet-trigger-based datasets is performed, where the first \pTavg{} bin of the single-jet-triggered result is replaced by the three lowest \pTavg{} bins from the pile-up-based approach.
This combined spectrum is re-fit using the NSC function, and the uncertainty on $N$ and $S$ is compared with the result fitting only the single-jet-trigger data as per Ref.~\cite{JETM-2018-05}.
The inclusion of the additional low-\pTavg{} bins from the pile-up dataset improves the statistical precision on $N$ by 40\%, and on $S$ by 20\%.

As a next step the 2017 and 2018 pile-up datasets are combined, and the NSC fit is performed to only the pile-up dataset.
The constant term in this fit is fixed to the value measured in Ref.~\cite{JETM-2018-05}, as the pile-up dataset does not have sensitivity to this high-energy part of the JER, while the noise and stochastic terms are left floating in the fit.
The resulting fit is shown in \Figure{fig:JER_final}, as compared with JER measured using a combination of the dijet asymmetry and random cone methods from Ref.~\cite{JETM-2018-05}.
Only statistical uncertainties are considered for pile-up JER extraction, while both statistical and systematic uncertainties on the combined JER are shown, as taken from Ref.~\cite{JETM-2018-05}.
The pile-up-based approach is shown to agree with the combined dijet and random cone fit: they agree within the uncertainty of the combined fit over the full range considered.
This result provides strong evidence that the pile-up dataset is indeed trigger-unbiased, as it was able to measure a physical quantity compatible with the result from single-jet-trigger-based approaches.
Moreover, it confirms that the pile-up dataset is an abundant source of low-energy dijet events, and thus demonstrates the potential of such an approach to extend the hadronic physics programme in new directions.

\begin{figure}[h!]
\centering
\subfigure[]{\includegraphics[width=0.48\textwidth]{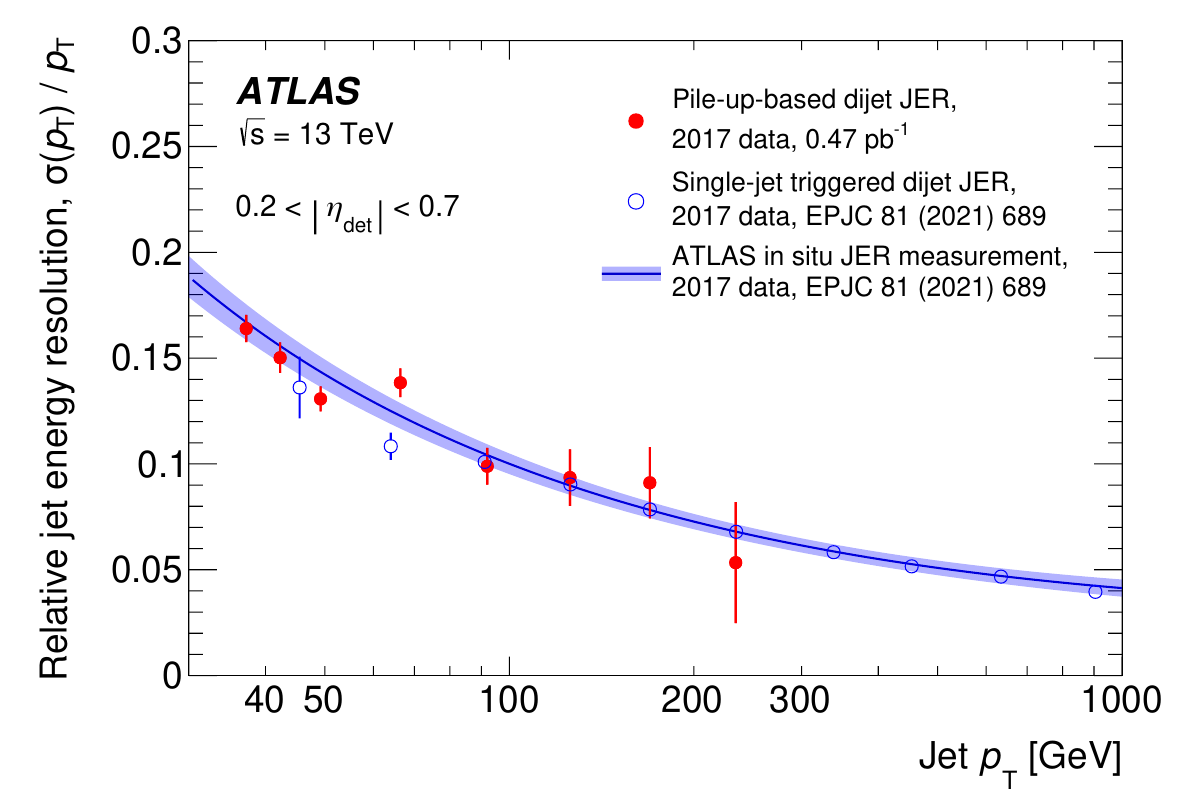}\label{fig:JER_dijet}}
\subfigure[]{\includegraphics[width=0.48\textwidth]{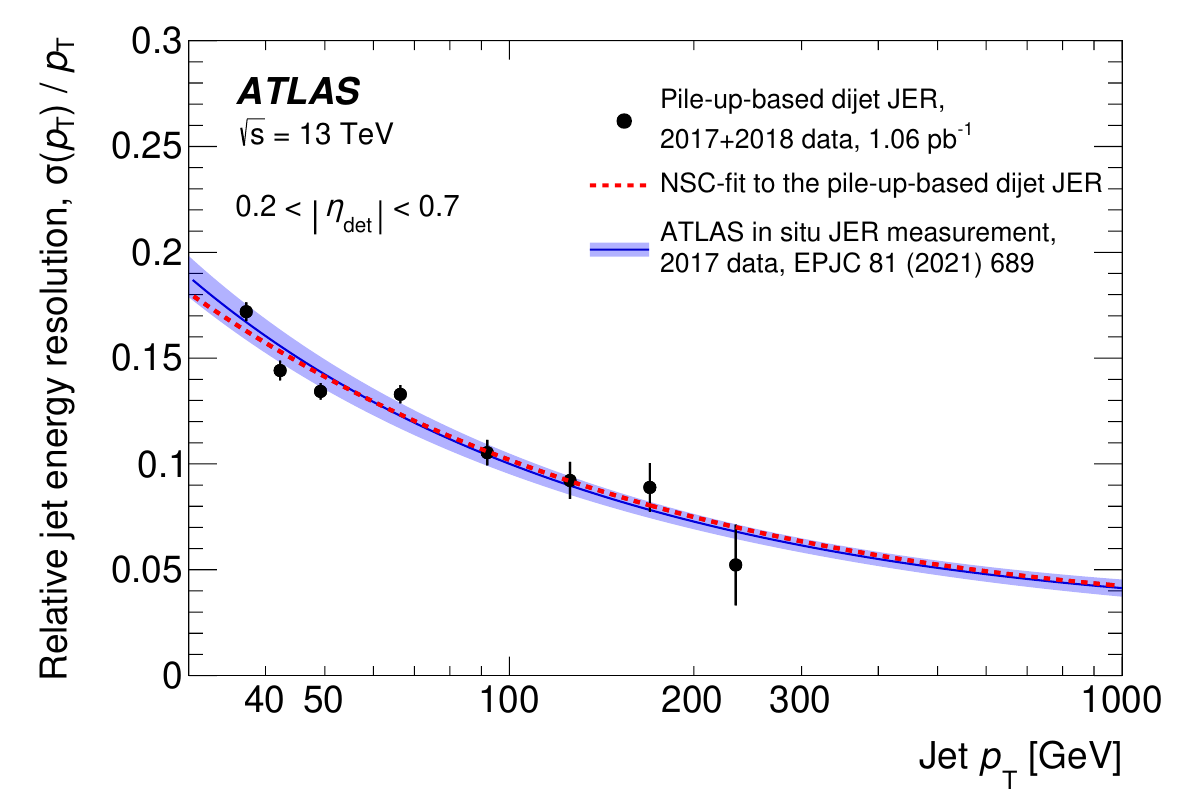}\label{fig:JER_final}}
\caption{
The JER, extracted from dijet events in the pile-up dataset, as compared with the (a) single-jet-triggered dijet JER and (b) combination of single-jet-triggered dijet JER and the random cone noise term results presented in Ref.~\cite{JETM-2018-05}.
The individual values for specific \pTavg{} bins are extracted from the dijet asymmetry.
The comparison with the single-jet-triggered dijet JER in (a) is performed using only the 2017 pile-up dataset, to better compare with the previous single-jet-trigger results, which also use only the 2017 dataset.
The NSC fit in (b) is performed on the 2017+2018 dataset to further improve the statistical precision, with the constant term fixed to the value measured in Ref.~\cite{JETM-2018-05}.
}
\label{fig:JER_extraction}
\FloatBarrier
\end{figure}


%
%
%
\FloatBarrier
\section{Conclusions}
\label{sec:conclusions}

%

The reconstruction of jets with respect to each primary vertex observed by the ATLAS detector represents a novel approach to using the LHC \ppp{} collision dataset, enabling the use of pile-up collisions for physics analyses, and providing access to an abundant source of low-energy hadronic physics processes.
The ATLAS 2015--2018 \ppp{} dataset was studied, corresponding to an integrated luminosity of up to 140\,\ifb{}, and with an average of 34 inelastic \ppp{} interactions per bunch crossing.
The integrated luminosity of the single-electron- and single-muon-triggered pile-up dataset was evaluated to be 0.47\,\ipb{} in 2017, 0.59\,\ipb{} in 2018, and 1.33\,\ipb{} for 2015--2018.
Comparisons with the corresponding single-jet-trigger integrated luminosities suggest that the pile-up dataset provides superior statistical precision for $\pT^\mathrm{lead}<60\,\GeV{}$, before considering analysis selection effects.
Various different possible biases and selection effects were studied and effectively mitigated, including aspects related to the ability to identify and remove the triggering collision and the handling of overlapping signals from different collisions.
As a demonstration of the potential of the dataset, the jet energy resolution is extracted from the balance of dijet events, and is found to be consistent with previous trigger-based results.
After including analysis selection effects, the pile-up dataset provides superior statistical precision for $\pTavg{}<65\,\GeV{}$ and up to 50 times more data than single-jet triggers for the jet energy resolution extraction at very low \pT{}, which is equivalent to a 240\,Hz single-jet trigger; such a trigger, if deployed during 2018, would have been the single highest rate trigger and would have accounted for roughly a fifth of the entire ATLAS trigger rate.
All together, this demonstrates a robust understanding of how pile-up collisions interact with the ATLAS detector, and showcases the viability and utility of the use of pile-up collisions as a trigger-unbiased dataset, thereby opening up new possibilities for low-energy hadronic physics data analysis at the LHC.


%

%
%
%
%
%
\section*{Acknowledgements}

%
%

%
%

We thank CERN for the very successful operation of the LHC and its injectors, as well as the support staff at
CERN and at our institutions worldwide without whom ATLAS could not be operated efficiently.

The crucial computing support from all WLCG partners is acknowledged gratefully, in particular from CERN, the ATLAS Tier-1 facilities at TRIUMF/SFU (Canada), NDGF (Denmark, Norway, Sweden), CC-IN2P3 (France), KIT/GridKA (Germany), INFN-CNAF (Italy), NL-T1 (Netherlands), PIC (Spain), RAL (UK) and BNL (USA), the Tier-2 facilities worldwide and large non-WLCG resource providers. Major contributors of computing resources are listed in Ref.~\cite{ATL-SOFT-PUB-2023-001}.

We gratefully acknowledge the support of ANPCyT, Argentina; YerPhI, Armenia; ARC, Australia; BMWFW and FWF, Austria; ANAS, Azerbaijan; CNPq and FAPESP, Brazil; NSERC, NRC and CFI, Canada; CERN; ANID, Chile; CAS, MOST and NSFC, China; Minciencias, Colombia; MEYS CR, Czech Republic; DNRF and DNSRC, Denmark; IN2P3-CNRS and CEA-DRF/IRFU, France; SRNSFG, Georgia; BMBF, HGF and MPG, Germany; GSRI, Greece; RGC and Hong Kong SAR, China; ISF and Benoziyo Center, Israel; INFN, Italy; MEXT and JSPS, Japan; CNRST, Morocco; NWO, Netherlands; RCN, Norway; MNiSW, Poland; FCT, Portugal; MNE/IFA, Romania; MSTDI, Serbia; MSSR, Slovakia; ARIS and MVZI, Slovenia; DSI/NRF, South Africa; MICIU/AEI, Spain; SRC and Wallenberg Foundation, Sweden; SERI, SNSF and Cantons of Bern and Geneva, Switzerland; NSTC, Taipei; TENMAK, T\"urkiye; STFC/UKRI, United Kingdom; DOE and NSF, United States of America.

Individual groups and members have received support from BCKDF, CANARIE, CRC and DRAC, Canada; CERN-CZ, FORTE and PRIMUS, Czech Republic; COST, ERC, ERDF, Horizon 2020, ICSC-NextGenerationEU and Marie Sk{\l}odowska-Curie Actions, European Union; Investissements d'Avenir Labex, Investissements d'Avenir Idex and ANR, France; DFG and AvH Foundation, Germany; Herakleitos, Thales and Aristeia programmes co-financed by EU-ESF and the Greek NSRF, Greece; BSF-NSF and MINERVA, Israel; NCN and NAWA, Poland; La Caixa Banking Foundation, CERCA Programme Generalitat de Catalunya and PROMETEO and GenT Programmes Generalitat Valenciana, Spain; G\"{o}ran Gustafssons Stiftelse, Sweden; The Royal Society and Leverhulme Trust, United Kingdom.

In addition, individual members wish to acknowledge support from Armenia: Yerevan Physics Institute (FAPERJ); CERN: European Organization for Nuclear Research (CERN PJAS); Chile: Agencia Nacional de Investigaci\'on y Desarrollo (FONDECYT 1230812, FONDECYT 1230987, FONDECYT 1240864); China: Chinese Ministry of Science and Technology (MOST-2023YFA1605700), National Natural Science Foundation of China (NSFC - 12175119, NSFC 12275265, NSFC-12075060); Czech Republic: Czech Science Foundation (GACR - 24-11373S), Ministry of Education Youth and Sports (FORTE CZ.02.01.01/00/22\_008/0004632), PRIMUS Research Programme (PRIMUS/21/SCI/017); EU: H2020 European Research Council (ERC - 101002463); European Union: European Research Council (ERC - 948254, ERC 101089007), Horizon 2020 Framework Programme (MUCCA - CHIST-ERA-19-XAI-00), European Union, Future Artificial Intelligence Research (FAIR-NextGenerationEU PE00000013), Italian Center for High Performance Computing, Big Data and Quantum Computing (ICSC, NextGenerationEU); France: Agence Nationale de la Recherche (ANR-20-CE31-0013, ANR-21-CE31-0013, ANR-21-CE31-0022, ANR-22-EDIR-0002), Investissements d'Avenir Labex (ANR-11-LABX-0012); Germany: Baden-Württemberg Stiftung (BW Stiftung-Postdoc Eliteprogramme), Deutsche Forschungsgemeinschaft (DFG - 469666862, DFG - CR 312/5-2); Italy: Istituto Nazionale di Fisica Nucleare (ICSC, NextGenerationEU), Ministero dell'Università e della Ricerca (PRIN - 20223N7F8K - PNRR M4.C2.1.1); Japan: Japan Society for the Promotion of Science (JSPS KAKENHI JP22H01227, JSPS KAKENHI JP22H04944, JSPS KAKENHI JP22KK0227, JSPS KAKENHI JP23KK0245); Netherlands: Netherlands Organisation for Scientific Research (NWO Veni 2020 - VI.Veni.202.179); Norway: Research Council of Norway (RCN-314472); Poland: Ministry of Science and Higher Education (IDUB AGH, POB8, D4 no 9722), Polish National Agency for Academic Exchange (PPN/PPO/2020/1/00002/U/00001), Polish National Science Centre (NCN 2021/42/E/ST2/00350, NCN OPUS nr 2022/47/B/ST2/03059, NCN UMO-2019/34/E/ST2/00393, UMO-2020/37/B/ST2/01043, UMO-2021/40/C/ST2/00187, UMO-2022/47/O/ST2/00148, UMO-2023/49/B/ST2/04085, UMO-2023/51/B/ST2/00920); Slovenia: Slovenian Research Agency (ARIS grant J1-3010); Spain: Generalitat Valenciana (Artemisa, FEDER, IDIFEDER/2018/048), Ministry of Science and Innovation (MCIN \& NextGenEU PCI2022-135018-2, MICIN \& FEDER PID2021-125273NB, RYC2019-028510-I, RYC2020-030254-I, RYC2021-031273-I, RYC2022-038164-I), PROMETEO and GenT Programmes Generalitat Valenciana (CIDEGENT/2019/027); Sweden: Swedish Research Council (Swedish Research Council 2023-04654, VR 2018-00482, VR 2022-03845, VR 2022-04683, VR 2023-03403, VR grant 2021-03651), Knut and Alice Wallenberg Foundation (KAW 2018.0157, KAW 2018.0458, KAW 2019.0447, KAW 2022.0358); Switzerland: Swiss National Science Foundation (SNSF - PCEFP2\_194658); United Kingdom: Leverhulme Trust (Leverhulme Trust RPG-2020-004), Royal Society (NIF-R1-231091); United States of America: U.S. Department of Energy (ECA DE-AC02-76SF00515), Neubauer Family Foundation.

%
%


%

%
%

%
%
%
%
%
%
%
%
%

%
%
%
\printbibliography
\clearpage
 
\begin{flushleft}
\hypersetup{urlcolor=black}
{\Large The ATLAS Collaboration}

\bigskip

\AtlasOrcid[0000-0002-6665-4934]{G.~Aad}$^\textrm{\scriptsize 104}$,
\AtlasOrcid[0000-0001-7616-1554]{E.~Aakvaag}$^\textrm{\scriptsize 17}$,
\AtlasOrcid[0000-0002-5888-2734]{B.~Abbott}$^\textrm{\scriptsize 123}$,
\AtlasOrcid[0000-0002-0287-5869]{S.~Abdelhameed}$^\textrm{\scriptsize 119a}$,
\AtlasOrcid[0000-0002-1002-1652]{K.~Abeling}$^\textrm{\scriptsize 56}$,
\AtlasOrcid[0000-0001-5763-2760]{N.J.~Abicht}$^\textrm{\scriptsize 50}$,
\AtlasOrcid[0000-0002-8496-9294]{S.H.~Abidi}$^\textrm{\scriptsize 30}$,
\AtlasOrcid[0009-0003-6578-220X]{M.~Aboelela}$^\textrm{\scriptsize 45}$,
\AtlasOrcid[0000-0002-9987-2292]{A.~Aboulhorma}$^\textrm{\scriptsize 36e}$,
\AtlasOrcid[0000-0001-5329-6640]{H.~Abramowicz}$^\textrm{\scriptsize 155}$
% ...
\bigskip
\\

% ...
$^{17}$Department for Physics and Technology, University of Bergen, Bergen; Norway.\\
% ...
$^{30}$Physics Department, Brookhaven National Laboratory, Upton NY; United States of America.\\
% ...
$^{36}$$^{(a)}$Facult\'e des Sciences Ain Chock, Universit\'e Hassan II de Casablanca;$^{(b)}$Facult\'{e} des Sciences, Universit\'{e} Ibn-Tofail, K\'{e}nitra;$^{(c)}$Facult\'e des Sciences Semlalia, Universit\'e Cadi Ayyad, LPHEA-Marrakech;$^{(d)}$LPMR, Facult\'e des Sciences, Universit\'e Mohamed Premier, Oujda;$^{(e)}$Facult\'e des sciences, Universit\'e Mohammed V, Rabat;$^{(f)}$Institute of Applied Physics, Mohammed VI Polytechnic University, Ben Guerir; Morocco.\\
% ...
$^{45}$Physics Department, Southern Methodist University, Dallas TX; United States of America.\\
% ...
$^{50}$Fakult\"{a}t Physik , Technische Universit{\"a}t Dortmund, Dortmund; Germany.\\
% ...
$^{56}$II. Physikalisches Institut, Georg-August-Universit\"{a}t G\"ottingen, G\"ottingen; Germany.\\
% ...
$^{104}$CPPM, Aix-Marseille Universit\'e, CNRS/IN2P3, Marseille; France.\\
% ...
$^{119}$$^{(a)}$New York University Abu Dhabi, Abu Dhabi;$^{(b)}$United Arab Emirates University, Al Ain; United Arab Emirates.\\
% ...
$^{123}$Homer L. Dodge Department of Physics and Astronomy, University of Oklahoma, Norman OK; United States of America.\\
% ...
$^{155}$Raymond and Beverly Sackler School of Physics and Astronomy, Tel Aviv University, Tel Aviv; Israel.\\
% ...
$^{a}$ Also Affiliated with an institute covered by a cooperation agreement with CERN.\\
$^{b}$ Also at An-Najah National University, Nablus; Palestine.\\
$^{c}$ Also at Borough of Manhattan Community College, City University of New York, New York NY; United States of America.\\
$^{d}$ Also at Center for Interdisciplinary Research and Innovation (CIRI-AUTH), Thessaloniki; Greece.\\
$^{e}$ Also at CERN, Geneva; Switzerland.\\
$^{f}$ Also at CMD-AC UNEC Research Center, Azerbaijan State University of Economics (UNEC); Azerbaijan.\\
% ...

\end{flushleft}


%

%
%

%

%
%
%
%

\end{document}